\documentclass{article}
\usepackage{amsmath}

\usepackage[figuresright]{rotating}

\usepackage{mathrsfs}
\usepackage{times}
\usepackage{bm}
\usepackage[numbers]{natbib}

\usepackage{graphics,color,pict2e}
\usepackage{epsfig}
\usepackage{xcolor}
\newtheorem{Theorem}{Theorem}

\def\bSig\mathbf{\Sigma}

 \newcommand{\ind}{\perp\!\!\!\!\perp}
\newcommand{\cdt}{\ensuremath{\cdot}}
\newcommand{\TE}{\ensuremath{T}}
\newcommand{\DE}{\ensuremath{D}}
\newcommand{\IE}{\ensuremath{I}}
\newcommand{\pr}{\ensuremath{\lambda}}
\newcommand{\prs}{\ensuremath{\lambda_s}}
\newcommand{\pra}{\ensuremath{\lambda_a}}

\renewcommand{\thefootnote}{\fnsymbol{footnote}}

\usepackage[figuresright]{rotating}

\begin{document}


\title{Mediation Analyses for the Effect of Antibodies in  Vaccination}

\author{MICHAEL P. FAY$^\ast$, DEAN A. Follmann \\[4pt]
\textit{Biostatistics Research Branch, National Institute of Allergy and Infectious Diseases,
Rockville, Maryland, USA}
\\[2pt]
{mfay@niaid.nih.gov}}

\markboth%
{M.P. Fay and D.A. Follmann}
{Mediation for Antibodies in Vaccination}

\maketitle

\renewcommand{\thefootnote}{\fnsymbol{footnote}}
\footnotetext{$^\ast$ To whom correspondence should be addressed.}
\renewcommand{\thefootnote}{\arabic{footnote}}

\begin{abstract}
{We review standard mediation assumptions as they apply to identifying antibody effects in a randomized vaccine trial and propose new study designs to
allow identification of an estimand that was previously unidentifiable.
For these mediation analyses, we partition the total ratio effect (one minus the vaccine effect) from a randomized vaccine trial into indirect
(effects through antibodies) and direct effects (other effects).
Identifying  $\pr$, the proportion of the total effect due to an indirect effect, depends on a cross-world quantity, the potential outcome among
vaccinated individuals with antibody levels as if given placebo, or vice versa.
We review assumptions for identifying $\pr$ and show that  there are two versions of $\pr$,
unless the effect of adding antibodies to the placebo arm is equal in magnitude to the effect of  subtracting antibodies from the vaccine arm.
We focus on the case when individuals in the placebo arm  are unlikely to have the needed antibodies. In that case, if a
standard assumption (given confounders, potential mediators and potential outcomes are independent) is true, only one version of $\pr$ is identifiable, and if not neither is identifiable.
We propose alternatives for identifying the other version of $\pr$, using experimental design to identify a formerly cross-world quantity.
Two alternative experimental designs use a three arm trial with the extra arm being passive immunization (administering monoclonal antibodies),
with or without closeout vaccination. Another alternative is to combine information from a placebo-controlled vaccine trial with a
placebo-controlled passive immunization trial.}

\end{abstract}

\maketitle

\section{Introduction}
\label{sec-intro}

Our goal is identifying the proportion of the vaccine effect from a randomized placebo-controlled vaccine trial that is due to the antibodies induced by the vaccine, say $\pr$.
In the mediation literature, the total effect is partitioned into an indirect effect that acts through the mediator (e.g., the vaccine effect that acts through the antibody response)
and a direct effect that acts directly on the outcome (e.g., the rest of the vaccine effect).\footnote{Unfortunately, in the vaccine literature, the term indirect vaccine effect
is used to describe another issue: the protective vaccine effect for a non-vaccinated individual due to nearby vaccinated individuals being less likely to be infectious,
see e.g., \citet{Hall:2010book}, Section 2.8. This paper is not about that type of indirect vaccine effect.}
Mediation analyses typically make certain positivity and independence assumptions in order to identify indirect and direct effects
(see \citet{Vand:2015} or Section~\ref{sec-sequential.ignorability}). This paper explores those assumptions in detail,
specifically focusing on the application to this antibody/vaccine problem.
For ease of exposition, unless stated otherwise, the term antibodies refers to only the particular type of antibody induced by the experimental vaccine.
Also for simplicity, we focus on the placebo-controlled vaccine trial, but many of the same issues may apply when another type of control arm is used instead of placebo (e.g., an arm that is given a control vaccine for another disease). Our focus is on the case when individuals in the placebo (or other control) arm do not have detectable levels of the antibody of interest, which is often the case
for new or rare diseases.

The antibody/vaccine mediation problem is important as seen by its application to COVID-19 vaccination. Large placebo-controlled vaccine trials showed impressive vaccine efficacies
during the stages of the pandemic before the omnicron variant was prevalent. It would be useful to be able to evaluate modifications of the vaccines
(e.g., dose changes, adjuvant changes, updating the antigen to current variants) under new populations and new exposures without having to re-run such
large trials. Instead, we could use antibodies to help predict vaccine efficacy. \citet{Gilb:2022} comprehensively explored the effects of antibodies on the vaccine
efficacy of the  Moderna (mRNA-1273) COVID-19 vaccine, including a mediation analysis using methods described in \citet{Benk:2021}.

Whereas \citet{Benk:2021} focuses on details of estimation and only briefly mentions standard mediation assumptions,  in this paper we focus on identifiability and more explicitly discuss the implications of the standard assumptions to the vaccine/antibody mediation problem.
The major difficultly of this identifiability problem is two cross-world quantities: the potential outcome of a vaccinee but with mediator values as if the vaccinee had gotten
placebo, or vice versa. These two quantities lead to two versions of indirect effects, which we call the {\it subtracting}
indirect effect (due to subtracting the antibody effect from the vaccinees)
or the {\it adding} indirect effect (due to adding antibody effect to the placebo participants).
(In the mediation literature these are called the {\it total} indirect effect and {\it pure} indirect effect, respectively \citep{Robi:1992}.)
We explore the typical mediation assumptions except without the often untenable assumption that the placebo participants have detectable antibodies specific to the infectious agent.
We show that the subtracting indirect effect and $\prs$ (its version of $\pr$) are identifiable, but the adding indirect effect and $\pra$ (its version of $\pr$) are not.
We further show that a key assumption for identifying the subtracting indirect effect is that
the potential outcomes are independent of the potential mediator responses conditional on confounders.
We show that in the simple case without confounders, the identification of $\prs$ by that assumption is equivalent to identifying a correlation of two potential
random variables as $0$ by assuming their independence.

Some major contributions of our paper are new ways to identify the adding indirect effect by supplementing the randomized  placebo-controlled vaccine trial with
extra experimental information.  This can be done by adding a randomized passive immunization arm to the randomized placebo-controlled vaccine trial,
where passive immunization is done by infusing (or injecting) individuals with monoclonal antibodies.
We discuss identifiability in that three arm trial, with and without a closeout vaccination in the passive immunization arm.
Alternatively, we can combine the placebo controlled vaccine trial with a second randomized trial of passive immunization versus placebo  to identify the adding indirect effect.
We first identify controlled vaccine effects from the placebo versus vaccine trial and then the controlled protective effects (which act like adding indirect effects)
from the placebo versus passive immunization trial. The latter effect does not require the conditional independence assumption between potential outcomes and mediators, since
that independence can be met by randomization to antibody values in the passive immunization arm.

Because this paper is partially about explaining  mediation analysis to vaccine researchers
and  vaccine trials to mediation experts,
we necessarily give extensive background on  the causal effects, vaccine biology, and  mediation definitions  (Section~\ref{sec-background}).
Section~\ref{sec-MedAssumptions} details typical mediation assumptions applied to this problem giving explicit identifiability results.
The binary mediator problem is addressed in Section~\ref{sec-binaryMediator} to give some intuition about the assumptions.
We describe identifying the adding indirect effect by supplementing the main placebo controlled vaccine trial with a passive immunization arm or experiment in
Section~\ref{sec-3arm.binary} (binary mediator case) and Section~\ref{sec-nonbinaryMediators} (general mediator case).

\section{Background}
\label{sec-background}

\subsection{Causal Vaccine Effects}
\label{sec-causalVE}

Vaccines work by exposing the vaccinee to an antigen, a protein on an outward facing part of the infectious agent.
The adaptive immune system of the vaccinee then produces activated antigen-specific B cells and helper T cells. Those cells
are part of the germinal center that creates more antigen-specific B cells, which in turn develop into either antibody-producing plasma cells or other types of cells such as memory B cells.
The plasma cells create antibodies specific to that antigen that travel to other areas  of the body through the blood. These antigen-specific antibodies
are usually detected from assays done on blood samples, and those assays are of different types
(e.g., an ELISA where the antibodies bind to proteins in a plate, or  a functional assay that measures how much the antibody neutralizes the agent).
In this paper, we will often refer to the mediator as ``antibodies'', and a more precise description would be the measured level of those antigen-specific antibodies
collected at a specific study time  from a blood sample using one assay. Although the antibodies measured in blood are one product of the vaccination,
some of those antigen-specific antibodies may be in other parts of the body (e.g., mucosal surfaces) and will not be detected by a blood sample and
hence would not be measured as the mediator.
Further,   there are many other responses to the vaccine that may lead to other mechanisms that may be helpful in protecting the vaccinee from developing
disease in response to a  future exposure from the infectious agent \citep[for more details, see e.g.,][]{Seig:2016}.

Vaccination is a way of preparing the immune system to be ready for future exposures. Vaccination is often especially helpful for individuals that have not previously been exposed, since those that have previously been exposed and survived will often have developed natural immunity. The gold standard for estimating vaccine efficacy is a
controlled randomized clinical trial. Prior to the availability of approved vaccines (e.g., for HIV, or in the early stages of the COVID-19 pandemic), the primary efficacy analysis from vaccine randomized trials typically exclude individuals that already have antibodies protective against the infectious agent of interest, since that population does not have the greatest need of vaccination. Further, because it takes time for the vaccinee to develop antibodies, vaccine trials also typically exclude individuals from the primary efficacy analysis data set that became diseased before the time when the peak antibody level is expected to be reached, usually 1 to 4 weeks after the last dose of the vaccination \citep[see e.g.,][Table 1]{Rapa:2022}.

\subsection{Vaccine Efficacy  Using Potential Outcomes}
\label{sec-VE.potential.outcomes}

We introduce formal causal notation, first for vaccine efficacy, then for mediation.
Because the paper will focus on identifiability not estimation, we do not need subscripts
to differentiate individuals in the population.
For example, we
let $A=1$ if the $i$th participant is randomized to vaccine, and $A=0$ if they are randomized to placebo, and
no  $i$ subscript is needed. Thus, $A$ represents the allocated arm of a typical (i.e., randomly selected) individual.
Let $Y=1$  if the typical participant gets the event (e.g., has confirmed disease)   during the time they are at risk during the study, and $Y=0$  if they do not.
Write the potential outcomes  for the typical participant as both $Y_0$, the outcome if they were randomized to placebo, and $Y_1$, the outcome if they were randomized to vaccine.
Although we can only observe one of the two potential outcomes per individual, we can compare the expected response in both arms, because by
randomization each individual has a positive and known probability of being in either of the arms.

Vaccine efficacy (VE) is defined as  1 minus a ratio effect (e.g., incidence ratio, hazard ratio, odds ratio)  \citep[see e.g.,][Table 2.2]{Hall:2010book},
and in this paper we define it as, $VE = $
$1- E(Y_1)/E(Y_0)$.
For vaccine mediation, sometimes odds ratios are used \citep[see e.g.,][]{Cowl:2019,Nguy:2015}.
Fortunately, when only a small percentage of the placeboes have the event, there is little difference between the different ratios \citep{Hudg:2004}.
The $VE$ estimand we use is reasonable in a randomized trial because the exposure processes are approximately balanced between the two arms \cite{Senn:2022}.
For example, although exposure may change by geographic area, the distributions of geographic areas by arm are balanced by randomization.
Similarly, although exposure will vary from individual to individual because of differential calendar times joining the study or censoring times leaving the study,
 those differences will be approximately equally distributed in the two arms as long as the rate of infection is not large, and the study entry times and censoring
 times are independent of the potential outcomes.

Unlike a difference effect (e.g., $E(Y_1)-E(Y_0)$), a ratio effect is relatively invariant to differences in exposure,
so it is useful for a vaccine trial since we can rarely predict well how many participants will be exposed
during the course of the trial.
To see this invariance, we reexpress the potential outcomes. Let $Y^*_a$ be the indicator of whether the typical participant would get the event
when they were in arm $a$ if they were exposed ($Y^*_a=1$ is yes, $Y^*_a=0$ is no).  Let $Z=1$ represent if they were exposed during the study (1=yes, 0=no),
so that $Y_a=Z Y^*_a$. If we assume the exposure is independent of whether they would be protected if exposed (which is a reasonable assumption in a blinded randomized trial), then
\begin{eqnarray}
VE & = & 1 - \frac{ E(Y_1)}{E(Y_0) } = 1 - \frac{ E(Z Y^*_1)}{E(Z Y^*_0) } = 1 - \frac{ E(Z) E( Y^*_1)}{E(Z) E( Y^*_0) } = 1 - \frac{ E( Y^*_1)}{E( Y^*_0) }. \label{eq:VE.ZYstar}
\end{eqnarray}
The vaccine efficacy does not depend on the exposure rate, but only on the differential in the potential to be protected given exposed, which is the scientific effect of interest.


\subsection{Defining Direct and Indirect Effects}
\label{sec-defining.DE.IE}

Now consider  mediation. Let $M_a$ be the potential mediator for a typical individual when $A=a$. In our example $M_1$ is the antibody level (e.g., IC50 titer at day 29 after vaccination) if  randomized to vaccine, and $M_0$ that level if randomized to placebo. Just as for the outcomes, we can only observe one of the potential antibody
levels for each individual. To explicitly denote the effect of the antibody on the outcome, we now write the outcome for the typical individual with $A=a$ as $Y_{aM_a}$.
More generally, we let $Y_{am}$
 be a potential outcome variable where $A$ is set to $a$ and $M$ is set to $m$ (and $m$ does not need to equal $M_a$).
This notation implies that only the values of $a$ and $m$ matter in the response, not the manner in which those values are assigned (e.g., if the mediator response for individual $i$ is $m$, it does not matter that the mediator value of $m$ occurred naturally in response to a vaccine or placebo (i.e., $M_a=m$), or was it set to $m$ by some other external intervention). This is known as the consistency assumption \citep{Cole:2009}, and we will assume that for now, but revisit it later.
Under consistency, the observed response, $Y$, is $Y = Y_{AM_A}$ and the observed mediator, $M$, is $M=M_A$.
In this notation,
\begin{eqnarray}
VE & = & 1-\theta_{\TE} = 1 - \frac{ E \left[ Y_{ 1M_1}  \right] }{E \left[ Y_{ 0M_0}  \right] },
\end{eqnarray}
where $\theta_{\TE}$ represents the total ratio effect.

Typically mediation effects are defined as differences \citep[see e.g.,][]{Hafe:2009}, but for vaccines
we define mediation effects in terms of ratio effects.
We partition the total ratio effect ($\theta_{\TE}$) into the product of an indirect ratio effect ($\theta_{\IE}$) and a direct ratio effect ($\theta_{\DE}$).
Let $\theta_{\IE}= \theta_{\TE}^{\pr}$ and $\theta_{\DE} = \theta_{\TE}^{(1-\pr)}$,
giving $\theta_{\TE} = \theta_{\IE} \theta_{\DE}$.
Thus, on the log scale $\log \left( \theta_{\TE} \right) = \log \left(  \theta_{\IE} \right) + \log \left(  \theta_{\DE} \right),$
 indirect and direct effects are additive, and $\pr$ is the proportion of the  total log-ratio effect due to indirect log-ratio effects.
By algebra, $\pr = \frac{ \log( \theta_{\IE} ) }{ \log( \theta_{\TE} ) }.$
The proportion mediated effect was defined this way in \citet{Gilb:2022} (see e.g., Table S9), although its explicit form and motivation was not given.

\begin{figure}

\includegraphics{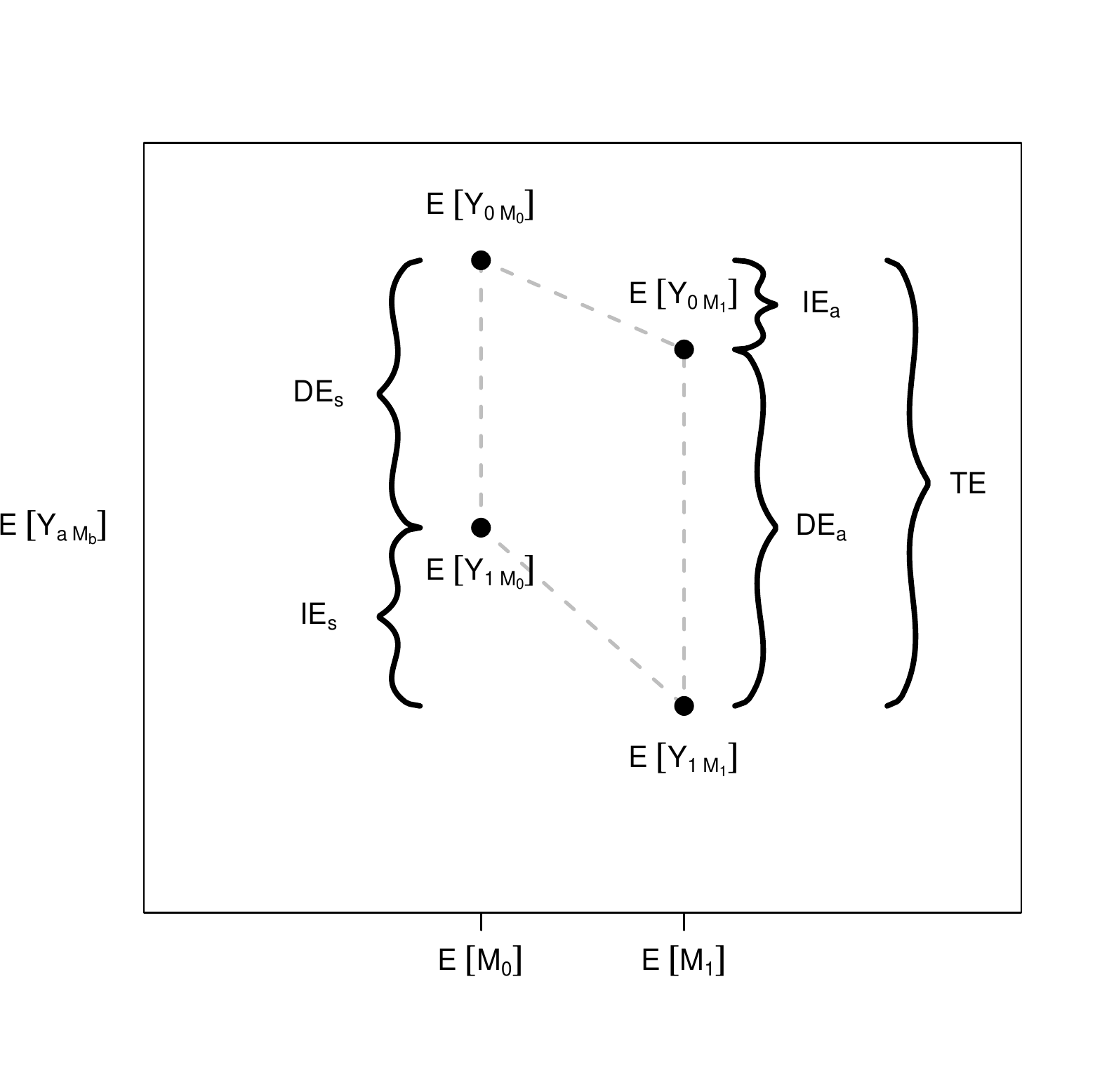}

\caption{Graphic of different ways of partitioning the total effect (TE) into an indirect effect (IE) and direct effect (DE). The vertical axis is on the log scale for ratio effects, and on the arithmetic scale for difference effects.
Moving southeast from  $E \left[ Y_{0M_0} \right]$ to
$E \left[ Y_{0M_1} \right]$ corresponds to adding in antibodies, while moving northwest from  $E \left[ Y_{1M_1} \right]$ to  $E \left[ Y_{1M_0} \right]$
corresponds to subtracting out antibodies. The Robins and Greenland \cite{Robi:1992} terminology is {\it pure} indirect effect ($IE_a$),
{\it total} direct effect ($DE_a$), {\it total} indirect effect ($IE_s$), and  {\it pure} direct effect ($DE_s$).
\label{fig-IEDEplot}}

\end{figure}

An indirect effect is the effect of changing only the mediator while holding the rest of the effect of vaccine constant. There are two main ways to define an indirect ratio effect.
The first way is $\theta_{\IE_a}  =   \frac{  E \left[ Y_{0M_1} \right] }{E \left[ Y_{0M_0} \right] }$,  which measures the effect on the placebo arm (denominator) of changing the mediator to the value it would have after vaccination but without actually vaccinating participants (numerator).
The `a' subscript denotes ``adding'' antibody to unvaccinated participants at the value they would have gotten if vaccinated (i.e., $M_1$).
The associated direct effect is part of the total effect that does not go through the mediator, $\theta_{\DE_a} \equiv \frac{\theta_{\TE}}{\theta_{\IE_a}} =  \frac{  E \left[ Y_{1M_1} \right] }{E \left[ Y_{0M_1} \right] }$.
Sections~\ref{sec-3arm.binary} and \ref{sec-nonbinaryMediators} focus on identification of $\theta_{\IE_a}$.
The second way to define an indirect effect is $\theta_{\IE_s}  =   \frac{  E \left[ Y_{1M_1} \right] }{E \left[ Y_{ 1M_0} \right] }$,
which measures the effect on the vaccine arm (numerator) of changing the mediator to the value it would have if the individual had been in the placebo arm (denominator).
The `s' subscript denotes ``subtracting'' the extra antibodies in the vaccinated participants, so that the antibody levels are what would have been seen in the placebo arm (i.e., $M_0$).
The associated direct effect is $\theta_{\DE_s} \equiv \frac{\theta_{\TE}}{\theta_{\IE_s}} =  \frac{  E \left[ Y_{1M_0} \right] }{E \left[ Y_{0M_0} \right] }$.
The mediation analysis in \citet{Gilb:2022} estimated $\theta_{\IE_s}$ and its associated $\pr$ value.
Figure~\ref{fig-IEDEplot} shows how  the total ratio effect can be partitioned these two ways,
\begin{eqnarray}
\theta_{\TE} & = & \theta_{\IE_a} \theta_{\DE_{a}} = \theta_{\IE_s} \theta_{\DE_s}. \label{eq:TEpartition1}
\end{eqnarray}
Let  $\pra$ be  $\pr$ under the first partition, and $\prs$ be its value under the second.

[Figure 1 about here.]

\citet{Robi:1992} used different terminology: $\theta_{\IE_a}$ is the {\it pure} indirect effect, $\theta_{\DE_a}$ is the {\it total } direct effect,
$\theta_{\IE_s}$ is the {\it total} indirect effect, and $\theta_{\DE_s}$ is the {\it pure} direct effect.
Alternatively to equation~\ref{eq:TEpartition1}, we can partition the total effect into 3 parts: pure indirect effect, the pure direct effect and an interaction effect (say, $\xi$),
$\theta_{\TE}  =  \theta_{\IE_a} \theta_{\DE_s} \xi,$
where
$\xi  =  \frac{E \left[ Y_{1M_1} \right] E \left[ Y_{0M_0} \right] }{ E \left[ Y_{1M_0} \right] E \left[ Y_{0M_1} \right]}.$
If $\xi=1$ then there is no interaction,  $\theta_{\IE_a}=\theta_{\IE_s}$ and $\theta_{\DE_{a}}=\theta_{\DE_s}$,
 the dotted quadrilateral in Figure~\ref{fig-IEDEplot} is a parallelogram, and all ways of partitioning the direct and indirect ratio effects are the same.
There are many other ways to define interaction \citep[see e.g.,][Section 7.6]{Vand:2015}.

\section{Mediation Analysis Using the Sequential Ignorability Assumptions}
\label{sec-MedAssumptions}

\subsection{Sequential Ignorability Assumptions}
\label{sec-sequential.ignorability}

In estimating direct or indirect effects, the difficult parameters to identify are either $E \left[ Y_{ 0M_1}  \right]$ (for identifying $\theta_{\IE_a}$) or $E \left[ Y_{ 1M_0}  \right]$ (for identifying $\theta_{\IE_s}$), because we observe no participant with those kinds of responses.
Those responses are called ``cross-world'' quantities because the intervention is in one world (e.g., where the participant was allocated to arm $A=a$) but the mediator is in another world (e.g.,
where the participant has mediator value as if allocated to arm $a'$,  $M_{a'}$,  where $a \neq a'$).
Thus, in order to identify $E \left[ Y_{ 1M_0}  \right]$ or $E \left[ Y_{ 0M_1}  \right]$ (and hence, to identify $\prs$ and $\pra$, respectively)  we need to make some assumptions that are not fully testable from the study data.
\citet{Pear:2001} showed that using a set of assumptions,  we can identify additive direct and indirect effects using a simple
equation that has become known as the mediation formula \citep[see e.g.,][]{Pear:2012}.
In this paper, since we deal with ratio effects, we focus on the piece of the mediation formula that estimates  $E \left[ Y_{ 1M_0}  \right]$ or $E \left[ Y_{ 0M_1}  \right]$ (equation~\ref{eq:EYaMb} in the following).
 \citet{Imai:2010} give two assumptions, which they called sequential ignorability assumptions, and those two assumptions are effectively a different statement of the assumptions of Pearl  \citep[see][p. 200-201]{Vand:2015}.
In this paper, we use those sequential ignorability assumptions, but for completeness we list the Pearl assumptions in Appendix~S1.
The situation the assumptions address is more general than the randomized trial situation, but it is important to review then since these set of assumptions are applied widely \citep[see e.g.,][]{Vand:2015,Nguy:2015},
including in vaccine applications (see e.g.,  \citet{Cowl:2019} and \citet{Gilb:2022})).

Let  $X$ be a vector of baseline covariates.
The sequential ignorability assumptions are:
\begin{quote}
{\it
\begin{description}
\item[$SI_1$:] $\left\{ Y_{am}, M_{a'} \right\} \ind A | X=x$,  and
\item[$SI_2$:]  $Y_{am} \ind  M_{a'} | A=a, X=x$,
\end{description}
for all $a,a'$ and $m$,
 assuming the following positivity assumptions,
\begin{description}
\item[$PosA$:] $Pr \left[ A=a| X=x \right]>0$ for all $a,x$,
\item[$PosM_0$:] $Pr \left[ M_0=m| A=0, X=x \right]>0$ for all $x$ and $m \in \mathcal{S}_{M}$, and
\item[$PosM_1$:] $Pr \left[ M_1=m| A=1, X=x \right]>0$ for all $x$ and $m \in \mathcal{S}_{M}$.
\end{description}
where $\mathcal{S}_{M}$ is the support for $M$.
}
\end{quote}
Often the ``sequential ignorability assumptions'' refers to not just $SI_1$ and $SI_2$, but also to the positivity assumptions, which are implicity assumed or listed as one positivity assumption \citep{Imai:2010}. In this paper it will be important to separate $SI_1$ and $SI_2$ (which we call the sequential ignorabiility assumptions) and the  different positivity assumptions. Later, as suggested by a referee, we explore modifications to the supports in the positivity assumptions.
Under the positivity assumptions listed above and the sequential ignorability  assumptions (and consistency),
the cross-world conditional expected potential outcomes are given by
\begin{eqnarray}
E \left[ Y_{aM_{a'}} | X=x \right]  & = & \sum_{m} E \left[ Y | A=a, M=m, X=x \right] Pr \left[ M=m | A=a', X=x \right]
 \label{eq:EYaMb}
\end{eqnarray}
\citep[see e.g.,][p. 465]{Vand:2015}.

\subsection{Application to the Vaccine/Antibody Mediation Analysis}
\label{sec-sequentialIgnorability.vaccine.Ab}

Now consider identifying $\prs$ or $\pra$ from a randomized placebo-controlled vaccine trial.
First, the assumptions $SI_1$ and $PosA$ are met for any randomized trial for any set of baseline variables $X$.
In this section, we discuss the harder assumptions to justify for our scenario, which  are consistency, $SI_2,$ $PosM_0,$ and $PosM_1$.

 Let the support of $M_0$ and $M_1$  be
$\mathcal{S}_M = \left\{ 0^*, \mathcal{S}_{detect} \right\}$, where $M=0^*$ represents $M$ below the limit of detection, and $M>0^*$ means  $M \in \mathcal{S}_{detect}$, and
$\mathcal{S}_{detect}$ is the set of possible detectable antibody values.
Consider the case of a new pathogen for which it is possible that  some individuals will have $M_0=0^*$.
Then, a violation of $PosM_1$ occurs  when the vaccine induces antibodies for the all individuals with $X=x$,
so that $Pr[ M_1=0^* | A=1, X=x] = 0$. A solution is to change the time of antibody measurement to
sooner after vaccination so that
there are some participants with $M_1=0^*$ within each $x$ and we can assume that $PosM_1$ is not violated, and then the resulting $\prs$ estimate can be used as a lower bound for the originally defined $\prs$ (i.e., the one using the original time of antibody measurement) under a reasonable monotonicity assumption \citep[see e.g.,][]{Gilb:2022}.
Changing the timing of the antibody measurement has a downside in that it may be less associated with the response, $Y_{AM_A}$,
since the original timing is typically designed to be at peak antibody levels.

Another problem, and a more common issue with vaccine trials, is violation of $PosM_0$ so that the analysis data set will have $Pr[ M_0>0^* | A=0, X=x] = 0$ for some $x$.
In many cases that probability is $0$ for all $x$.
For example, we could have all $M_0=0^*$ for a new infectious agent, because no one would have been previously exposed.
Another example is if we exclude from the study those who have antibody detectable at baseline
as well as those exposed to the infectious agent before the time of measurement of the antibody (i.e., study time when $M$ is measured).
Consider equation~\ref{eq:EYaMb} with $a=0$ and $a'=1$, giving the cross-world mean representing adding antibody to the unvaccinated,
\begin{eqnarray}
E \left[ Y_{0M_{1}} | X=x \right]  & = & \sum_{m} E \left[ Y | A=0, M=m, X=x \right] Pr \left[ M=m | A=1, X=x \right].  \label{eq:EY0M1}
\end{eqnarray}
The expression $E \left[ Y | A=0, M_0=m, X=x \right]$ for $m>0^*$ in equation~\ref{eq:EY0M1}
is not identifiable
when $Pr[ M_0>0^* | A=0, X=x]=0$, because in that case we do not observe $Y$ given $A=0$ and $M_0=m$ for any $m>0^*$
on any individual.
When  $a=1$ and $a'=0$,  both sequential ignorability assumptions hold, and all positivity assumptions hold except for allowing $Pr[ M_0>0^* | A=0, X=x]=0$
(i.e., no detectable antibody in the placebo arm), then equation~\ref{eq:EYaMb} can still be used (see Supplement Section~S2)
to give
\begin{eqnarray}
E \left[ Y_{1M_{0}} | X=x \right]  & = &   E \left[ Y | A=1, M=0^*, X=x \right].  \label{eq:EY1M0}
\end{eqnarray}
The right-hand side of equation~\ref{eq:EY1M0}  is just the disease rate among vaccinated individuals with undetectable antibody among study participants with baseline $X=x$.
Using equation~\ref{eq:EY1M0}, we can identify $E \left[ Y_{1M_{0}}  \right]$
as
\begin{eqnarray}
{E} \left[ Y_{1M_{0}}  \right]  & = &  \sum_x  {E} \left[ Y | A=1, M=0^*, X=x \right] Pr \left[ X=x \right],  \label{eq:EY1M0sumx}
\end{eqnarray}
where the summation is over the possible values of $X$,
and ${E} \left[ Y | A=1, M=0^*, X=x \right]$ is estimated with the sample mean of $Y$ among the vaccinated with undetectable antibody responses and $X=x$.


Consider the consistency assumption \citep{Cole:2009} in this antibody/vaccine mediation scenario.
Suppose we define the antibody mediator as the amount of antibody measured at a particular time post vaccination, and  an indirect effect is through that antibody mediator and the rest of the total effect is a direct effect. Consider the plasma cells that produce antibodies which give the antibody mediated vaccine effects. After the antibody mediator is measured, the plasma cells  (especially if they are long-lived plasma cells \citep[see][]{Seig:2016}) may play a part in protection by producing more antigen-specific antibodies in the future. By the strict definition, those latter antibodies would be seen as direct effects.
 If we wish to apply the mediation analysis results to
predict the proportion of total vaccine effect due to that antibody mediator from a similar vaccine, then
the consistency assumption may approximately hold because the second similar vaccine will likely also produce plasma cells that remain after the antibody is measured.
In contrast, if we wish to apply the mediation results from the original vaccine to predict the proportion of the total effect due to antibodies in the case when
the intervention is infusion with monoclonal antibodies, then the consistency assumption will likely be violated (i.e., the two mediation analyses will be estimating different effects).
It is likely that an individual's outcome under monoclonal antibodies  will be different than if they had an antibody response due to a vaccine, because the vaccine will induce plasma cells
which may continue to make antibodies, and hence have a larger direct effect.
This violation may not be a problem if the amount of antibodies detected  produces overwhelming protection and additional antibodies would not affect risk much.
Even if the consistency assumption was violated, the mediation analysis might be useful to estimate bounds on an indirect effect
if we could assume that the vaccine induced antibodies are at least as effective as the addition of externally produced monoclonal antibodies.

Next, consider the $SI_2$ assumption.
For this we once again simplify the immunology. Suppose we can partition the antigen-specific antibody response into two parts:
the creation of the antibodies, and the subsequent protective function of the antibodies.
Some protective functions of the antibodies are \citep[][Table 2.1]{Seig:2016}:
(i) binding to toxins produced by the infectious agent to stop their deleterious effect,
(ii) stopping replication of a virus  by binding to the virus and preventing the virus from entering cells of the host,
(iii) opsonizing, where the antibodies mark pathogens outside cells so that other immune cells (e.g., macrophages or neutrophils) may clear them,
(iv) activating  other processes (e.g., the complement cascade).
The $SI_2$ assumption (i.e., within each arm, conditional on baseline covariates, $X$, the   potential antibody response, $M_{a'}$,
is independent on the  potential outcomes for the study,
$Y_{am}$, for all $a,a',$ and $m$),
implies that within levels of $X$, the processes that create antibodies are independent from the processes that
lead to their protective functions. Because the body has evolved to have a useful immune system, it is possible that there would be a dependence between those two processes; however,
there may not be a high dependence (e.g., \citet{Goel:2021} showed that antibody level post-boost did not have any substantial correlation with
post-boost memory B-cells, the latter of which are suspected to be related to protection).
If there is such a dependence, in order to meet assumption $SI_2$, we would need to find baseline covariates, $X$,  such that conditional on $X$, the
potential antibody response vector is independent of the potential outcome vector. For example, suppose the baseline covariates can be used to classify individuals
into $k$ immune classes, where some
 classes have healthier immune systems than others. If {\it within each class} we can assume $Y_{am} \ind M_{a'}$, then $SI_2$ would be met.
Consider an example with only 2 classes, healthy and immunocompromised,
and suppose $X$ could be used to classify each individual into those two types. Further suppose that within both the healthy and the immunocompromised types
the process for creating antibodies is independent from the protective processes of the antibodies. Under those suppositions, $SI_2$ would be met.

In summary, we have detailed the main assumptions used in most, if not all, vaccine mediation analyses up to this point,
including a recent important vaccine mediation analysis \citep[see e.g.,][]{Gilb:2022,Benk:2021}.
We have explicitly shown that when $PosM_0$ fails such that $Pr[ M_0>0^* | A=0] = 0$, then we can still identify the subtracting indirect effect ((i.e., total indirect effect),
as long as the randomization is valid and the $SI_2$ and $PosM_1$ assumptions are met.
We have focused on binary potential outcomes instead of time-to-event outcomes to avoid extra complications such as the timing of the outcomes and  censoring,
since those extra complications are peripheral to our main focus on identifiability.

\section{Binary Mediators and Responses}
\label{sec-binaryMediator}

From the previous section we showed that a key difficult assumption  is $SI_2$. In this Section, we study the simple case when the mediators are binary
to gain intuition about how $SI_2$ is affecting the identifiability of the estimators of $\pr$.
In this section,
let $M_A=1$ if $m \geq \tau$, and $M_A=0$ if $m< \tau$, where $\tau$ is some specified threshold.
The threshold $\tau$, need not be the limit of detection, it could be some positive value of $m$ related to the
antibody activity.
To keep the exposition simple, we will ignore baseline covariates,
but it is straightforward (although notationally cumbersome) to apply the models
of this section separately to each level of the baseline covariates if the analogous conditional independence assumptions are appropriate.

\subsection{Base Model}
\label{sec-base}

To start we gather together some assumptions that appear reasonable for the placebo-controlled randomized vaccine trial.
As a shorthand, we call this set of assumptions the ``base model''.
\citet{Robi:1992} studied the mediator problem with binary mediators and binary potential outcomes. Under this model there are  $2^6 = 64$ different types of possible vectors of potential mediator responses and potential outcomes (2 levels for the mediator,
and 4 levels for the paired outcomes).
\citet{Robi:1992} made three assumptions that reduce the 64 types to 18,   which we apply and translate to our vaccine/antibody example:
\begin{description}
\item [$RG_{1}$: ] Vaccination cannot block antibodies that would have been present under the placebo arm. This means we never have types where $M_1=0$ and $M_0=1$.
\item[ $RG_{2}$:]  Vaccination cannot cause the disease. This means we never have types where $Y_{1m}=1$ and $Y_{0m}=0$.
\item[ $RG_{3}$:]   Antibodies cannot cause the disease. This means we never have types where $Y_{a1}=1$ and $Y_{a0}=0$ for any $a$.
\end{description}
These assumptions are often called monotonicity assumptions \citep[see e.g.,][]{Hafe:2011}.
These seem reasonable assumptions for many vaccines, although there is one known case of a violation of $RG_2$ (a vaccine for Dengue virus for one type enhancing the
probability of disease in another type \citep[see e.g.,][]{Shuk:2020}).
Additionally, $RG_2$ excludes cases where vaccination increases the risk of disease.
For example, some individuals may change their behaviour to increase their risk of infection if they suspect they have been vaccinated and believe that the vaccination works.
If that behaviour change results in a disease case, that would count as the vaccination causing the disease.
Because of this, in vaccine trials double-blinding is especially important for justifying $RG_2$. Related to that, the reactogenicity of the vaccine may unintentionally unblind some participants. To alleviate that problem,  sometimes the control arm uses a vaccine for a different disease than that of the experimental vaccine.

For the vaccine example, because we exclude individuals that had antibodies at baseline, and because we exclude anyone with detectable disease before the antibody is measured, it is very unlikely that anyone left in the analysis data set in the placebo arm will have a positive antibody response. So we assume that we never have types where $M_0=1$. This reduces the types of interest to 12.

These types are listed in Table~\ref{tab-MY}, with each type described in the last column.
The first column gives the notation for the proportion of the population with potential outcomes ${\bf Y}=[Y_{11},Y_{10},Y_{01},Y_{00}]$ and potential mediators ${\bf M}=[M_1, M_0]$,
which is  $\pi^{\bf M}_{\bf Y}$. For example, $\pi^{10}_{0101}$ is the proportion with ${\bf M}=[10]$ (i.e., $M_1=1$ and $M_0=0$)
and ${\bf Y}=[0101]$ (i.e., $Y_{11}=0, Y_{10}=1, Y_{01}=0$ and $Y_{00}=1$). For a person with that type, $Y_{10}$ and $Y_{01}$ are cross-world potential outcomes.

[Table~\ref{tab-MY} about here]

Under a randomized trial, $A$ is independent of both the potential outcome vector (i.e.,  $A \ind {\bf Y}$)
and the potential mediator response vector
(i.e., $A  \ind {\bf M}$).
Altogether, the assumptions for the {\it base} model described in this section are $RG_1$, $RG_2$, $RG_3$, $Pr[ M_0=1]=0$, $A \ind {\bf Y}$, and $A \ind {\bf M}$.
We feel that  often these assumptions will be quite reasonable.

To find identifiable parameters, we first re-express the 12 proportions of Table~\ref{tab-MY}
as Table~\ref{tab-strataDefns}, which lists the 12 proportions in a $4 \times 2$ table with observable margins,
and a $4 \times 6$ table with one of the observable columns partitioned into 5 sub-columns. The 6 observable (and hence identifiable) margins are denoted with $\phi_{amy}$ parameters,
where the $amy$ subscripts represent respectively, $A$
($v=$vaccine, $p=$placebo),
$M_A$ ($a=$antibody response positive, $n=$no antibody response), and  $Y_{AM_A}$ ($f=$ failure [had disease],  $s=$ success [did not have disease]).
For example, $\phi_{vaf}$ is the proportion of the population that if randomized to {\bf v}accine would produce {\bf a}ntibodies and have an event (i.e., would {\bf f}ail).
Recall, in the base model $Pr[M_0=1]=0$ so  $\phi_{pas}=0$ and $\phi_{paf}=0$ and are not listed in Table~\ref{tab-strataDefns}.

[Table 2 about here.]

To be identifiable under the base model is to be able to express a parameter in terms of the $\phi$ parameters.
From Table~\ref{tab-strataDefns} we see that
both $E(Y_{1M_1})$ and
$E(Y_{0M_0})$ are  identifiable,
\begin{eqnarray}
E \left[ Y_{1M_1} \right] & = & \phi_{vaf} + \phi_{vnf}  \nonumber \\
E \left[ Y_{0M_0} \right] &= & \phi_{pnf}. \nonumber
\end{eqnarray}
On the other hand, the cross-world expectations
are not identifiable without further assumptions, so are given with a combination of identifiable
and other parameters (see Tables~\ref{tab-MY} and \ref{tab-strataDefns}),
\begin{eqnarray}
E \left[ Y_{1M_0} \right]  = &  \phi_{vaf} + \phi_{vnf} + \pi_{0101}^{10} + \pi_{0111}^{10}
  & =  E \left[ Y_{1M_1} \right] + \pi_{0101}^{10} + \pi_{0111}^{10}  \label{eq:EY1M0.eq.EY1M1+e3+e4} \\
E \left[ Y_{0M_1} \right]  = &  \phi_{vaf} + \phi_{vnf} + \pi^{00}_{00 \cdot 1} + \pi_{0011}^{10} + \pi_{0111}^{10}
&  =  E \left[ Y_{1M_1} \right] + \pi^{00}_{00 \cdot 1} + \pi_{0011}^{10} + \pi_{0111}^{10},  \label{eq:EY0M1.eq.EY1M1+b+e2+e4}
\end{eqnarray}
where $\pi^{00}_{00 \cdot 1} = \pi_{0001}^{00} + \pi_{0011}^{00}$,
and the ``$\cdot$'' denotes
summations over an index.
From equation~\ref{eq:EY1M0.eq.EY1M1+e3+e4}, we see that to identify
$E \left[ Y_{1M_0} \right]$, we need to be able to identify $\pi_{0101}^{10} + \pi_{0111}^{10}$.
 By inspection of the definition of the $\phi$ parameters in Table~\ref{tab-strataDefns}, we see that no combination of the $\phi$ parameters
are able to identify  $\pi_{0101}^{10}+ \pi_{0111}^{10}$ (see third row, second column).
Therefore, $E \left[ Y_{1M_0} \right]$ is not identifiable from the base model. Further, under the base model the proportion of the total effect due to an indirect effect, $\pr$, is not identifiable.
We formally show this in Theorem~\ref{Theorem.prsBounds}.

\begin{Theorem}
\label{Theorem.prsBounds}
Any
 $\prs \in \left[ 0,1 \right]$ is compatible with the base model.   Similarly, any
 $\pra \in \left[ 0,1 \right]$ is compatible with the base model.
\end{Theorem}
The proof is in Supplement Section~S3.
Theorem~\ref{Theorem.prsBounds} shows that the base model is inadequate to
get any information about $\pr$ (either $\prs$ or $\pra$) from the data; more assumptions are needed when $\pr$ is the interest.

\subsection{Model with Potential Mediators Independent of Potential Outcomes}

We previously discussed in Section~\ref{sec-MedAssumptions} the sequential ignorability assumptions,
and how they may be applied even when $Pr[ M_0=1| A=0]=0$ to estimate $E \left[ Y_{1 M_0} \right]$ (see equation~\ref{eq:EY1M0sumx} which additionally adjusts for baseline covariates).
The base model essentially already includes $SI_1$, and now we explore adding
 $SI_2$  to the base model. In Section~\ref{sec-sequentialIgnorability.vaccine.Ab} we discussed why it is difficult to justify $SI_2$ in the vaccine/antibody mediation analysis; nevertheless,
because the sequentially ignorability assumptions are common and allow some identifiability,  we  discus them here.
In this section (Section~\ref{sec-binaryMediator}) we have simplified the exposition,
leaving off the baseline variables, and  write the independence expression in $SI_2$ without explicitly conditioning on $A=a,X=x$ as  $Y_{am}  \ind M_{a'}$; it means that each mediator potential response  is independent of each potential outcome.
Let  $\mathcal{M}_{2}$ represent the model defined as  the base model with the added assumption from $SI_2$  that $Y_{am}  \ind M_{a'}$.

\begin{Theorem}
\label{Theorem.MindY}
Under $\mathcal{M}_{2}$,  then
\begin{description}
\item[(i)] $\pi_{0000}^{00},$
$\pi_{00 \cdot 1}^{00}$, $\pi_{\cdot 1 \cdot 1}^{00}$, $\pi_{0000}^{10}$, $\pi_{0 \cdot \cdot 1}^{10}$, and $\pi_{1111}^{10}$ are identifiable, with
$\pi_{0000}^{00} = \phi_{pns} \phi_{vn}$,
$\pi_{00 \cdot 1}^{00} = \phi_{vns} - \phi_{pns} \phi_{vn}$,
$\pi_{\cdot 1 \cdot 1}^{00}=\phi_{vnf}$,
$\pi_{0000}^{10} = \phi_{pns} \phi_{va}$,
$\pi_{0\cdot \cdot 1}^{00} = \phi_{vas} - \phi_{pns} \phi_{va}$, and
$\pi_{1111}^{10}=\phi_{vaf}$, and $\phi_{va}=1-\phi_{vn} = \phi_{vaf}+\phi_{vas}$.
\item[(ii)] $E \left[ Y_{0M_1} \right]$ is not identifiable, and
\item[(iii)] $E \left[ Y_{1M_0} \right]$ is identifiable and equal to $\frac{ \phi_{vnf} }{ \phi_{vn}  }$.
\end{description}
\end{Theorem}

%

The proof of Theorem~\ref{Theorem.MindY} is in Supplement Section~S4.
First, note that Theorem~\ref{Theorem.MindY}(i) allows us to test for certain violations of the assumptions,
because $\pi_{00 \cdot 1}^{00}$ and $\pi_{0\cdot \cdot 1}^{00}$ must be positive, or rewriting
\begin{eqnarray*}
\pi_{00 \cdot 1}^{00} & \geq & 0  \Leftrightarrow  \frac{ \phi_{vns} }{\phi_{vn} } \geq \phi_{pns}  \Leftrightarrow  \frac{ \phi_{vnf} }{\phi_{vn} } \leq \phi_{pnf} \\
\pi_{0\cdot \cdot 1}^{00} & \geq & 0  \Leftrightarrow  \frac{ \phi_{vas} }{\phi_{va} } \geq \phi_{pns}  \Leftrightarrow  \frac{ \phi_{vaf} }{\phi_{va} } \leq \phi_{pnf}.
\end{eqnarray*}
Second,  Theorem~\ref{Theorem.MindY}(ii) shows $E \left[ Y_{0M_1} \right]$ is not identifiable, which may seem strange because the usual positivity and sequential ignorability assumptions show identifiability (see equation~\ref{eq:EYaMb});
however, recall that $Pr[ M_0=1]=0$ from the base model, which violates the $PosM_0$ positivity assumption (see Section~\ref{sec-sequential.ignorability}). A consequence is that $\pra$ is not identifiable.
Finally, plugging in the value of $E \left[ Y_{1M_0} \right]$ from Theorem~\ref{Theorem.MindY}(iii),
we get $\theta_{\IE_s}  = $ $\frac{ E \left[ Y_{1M_1} \right] }{ E \left[ Y_{1M_0} \right] } = $ $\frac{ \phi_{vf} \phi_{vn} }{ \phi_{vnf} }$, so that
\begin{eqnarray}
\prs & = & \frac{ \log \left( \frac{ \phi_{vf} \phi_{vn} }{ \phi_{vnf} } \right)  }{ \log \left( \frac{ \phi_{vf}  }{ \phi_{pnf} } \right)}.
\label{eq:prs.MindY}
\end{eqnarray}
Thus, assuming $Y_{am}  \ind M_{a'}$ with the base model
determines $\prs$. This is a red flag, because we do not want the parameter of interest  to be identified entirely
by an assumption that is not strongly justified by the subject matter. Below we express different consequences of $SI_2$ to better critique its plausibility.

\begin{Theorem}
\label{Theorem.EquivProblems}
The following three statements are equivalent:
\begin{description}
\item[{\em Statement A}:] Under the base model, $E(Y_{1 M_0})$ ranges from $E(Y_{1M_1})$ to $E(Y_{0M_0})$, and  if we additionally assume $Y_{am}  \ind M_{a'}$, then $E(Y_{1 M_0})= \phi_{vnf}/\phi_{vf}$.
\item[{\em Statement B}:] Under the base model, $\prs$ ranges from $0$ to $1$, and  if we additionally assume $Y_{am}  \ind M_{a'}$, then $\prs$ is given by equation~\ref{eq:prs.MindY}.
\item[{\em Statement C}:] Under the base model, $\rho \equiv \mathrm{Corr}(Y_{1M_0}, M_1)$ ranges from $\rho_{min}$ to $\rho_{max}$, and if we additionally assume  $Y_{am}  \ind M_{a'}$ (which implies the assumption that $Y_{1M_0} \ind M_1$),
then $\rho=0$, where
\begin{eqnarray*}
\rho_{min} & = &   \frac{ \left\{  \phi_{pnf} - \frac{\phi_{vnf}}{\phi_{vn}} \right\} \phi_{vn} }{
 \sqrt{ (1- \phi_{vn} ) \phi_{vn}  \phi_{pnf} \left\{ 1 - \phi_{pnf} \right\} }} \\
\rho_{max} & = &  \frac{ \left\{  \phi_{vf} - \frac{\phi_{vnf}}{\phi_{vn}} \right\} \phi_{vn} }{
 \sqrt{ (1- \phi_{vn} ) \phi_{vn}  \phi_{vf} \left\{ 1 - \phi_{vf} \right\} }},
\end{eqnarray*}
\end{description}
 $\phi_{vn}=\phi_{vns}+\phi_{vnf}$, and $\phi_{vf}=\phi_{vnf}+\phi_{vaf}$.
\end{Theorem}

The theorem is proven in Supplement Section~S5.
Statement A restates Theorem 2(iii) by rewriting the conditions implied by the base model that require $\theta_T \leq \theta_{I_s} \leq 1$, i.e.,
that the vaccine cannot cause the event and the subtracting indirect effect cannot be harmful and is not as extreme as the total ratio effect.
The usual approach to solving this problem is Statement B,
which is just expressing $\prs$ as a function of $E(Y_{1M_1})$, $E(Y_{1 M_0})$, and $E(Y_{0M_0})$.
Another approach is Statement C, which is expressing $\rho=\mathrm{Corr}(Y_{1M_0}, M_1)$ as a function of $E(Y_{1 M_0})$ and some identifiable parameters, and the range for $\rho$ required by the base model just plugs in $E(Y_{1M_1})$ and $E(Y_{0M_0})$ into the expression for $\rho$.
The main point of Theorem~3 is that the usual approach to solving this problem (Statement~B) is like trying to identify a correlation,
and assuming independence of the two random variables to conclude that the correlation is $0$ (Statement C). Statement~C seems like assuming what we are trying to identify because independence and correlation are inherently linked,
 but our objective is estimating $\prs$
which is really $\prs(\rho)$, a complex function of $\rho$. The function $\prs(\rho)$  may be steep or relatively flat for $\rho$'s of interest, and the steepness
depends on  a given dataset.  Also, $\prs(0)$ depends on identifiable parameters specific to each setting (see equation~\ref{eq:prs.MindY}).

The models of this section ignore adjustments for baseline covariates. If those baseline covariates are measured and  control for all of the differences between the
antibody production processes and the antibody protection process, then  $SI_2$
may hold and
equation~\ref{eq:EY1M0sumx} is justified.
Those are the assumptions used in the mediation analysis for the Moderna (mRNA-1273) COVID-19 vaccine \citep{Gilb:2022,Benk:2021}.
If only some of those baseline covariates are used but are assumed to be all of them, the resulting estimator may be improved.
A problem is that it is often difficult to measure all variables needed to control for these processes and to know that
$SI_2$
approximately holds.

\subsection{Example}
\label{sec-example}

In Table~\ref{tab:toy} we show   the results of a made-up randomized placebo-controlled vaccine trial with 10,000 participants in each arm.
To keep it simple, we assume there are no baseline variables measured.
We estimate VE as 90\% since
$\hat{\theta}_{\TE} = \frac{ \hat{E} \left[ Y \left\{ 1,M(1) \right\} \right] }{ \hat{E} \left[ Y \left\{ 0,M(0) \right\} \right]} = $ $ \frac{10/10000}{100/10000} = 0.10$.
If  the sequential ignorability assumptions are met and $Pr[M_0=1| A=0]=0$,
then we can use equation~\ref{eq:EY1M0} with only 1 level of $X$ (which is equivalent to the result of Theorem~\ref{Theorem.MindY}), to get,
$\hat{E} \left[ Y_{1M_0}  \right]  =   \hat{E} \left[ Y_{1M_1}  | M_1=0 \right]   =  0.4\%$.
This gives us
$\hat{\theta}_{\IE_s}  = $ $ \frac{  \hat{E} \left[ Y \left\{ 1,M(1) \right\} \right] }{ \hat{E} \left[ Y \left\{ 1,M(0) \right\} \right] } = $ $ \frac{0.1\%}{0.4\%} = $ $ 0.25,$ and
$\hat{\theta}_{\DE_s}  = $ $\frac{  \hat{E} \left[ Y \left\{ 1,M(0) \right\} \right] }{ \hat{E} \left[ Y \left\{ 0,M(0) \right\} \right] } = $ $ \frac{0.4\%}{1\%} = $ $   0.40,$
and
$\hat{\lambda}_s  = $ $\frac{ \log \left( \hat{\theta}_{\IE_s} \right)  }{ \log \left(  \hat{\theta}_{\TE} \right) } = $ $0.6021,$
so that under model $\mathcal{M}_{2}$ we estimate that  about $60.21\%$  of the total ratio effect is mediated through the antibodies.

[Table 3 about here.]




\subsection{Three Arm Trial}
\label{sec-3arm.binary}

Of the models studied, only model $\mathcal{M}_{2}$  achieves identifiability of $\prs$, and no model gives identifiability of $\pra$. Thus, we consider an experimental solution.
Consider trial with three arms: vaccine, placebo, and passive immunization,
where passive immunization is infusing (or injecting) participants with antibodies made externally, such as monoclonal antibodies.
 We start with the assumptions of the base model of Section~\ref{sec-base} and add the third arm.
We make a consistency assumption about the antibodies, assuming that the antibody effects from infusion of monoclonal antibodies will be the same as if those antibodies
occurred due to vaccination.

We treat that intervention ($A=2$) as if it is a placebo with $M=1$, so we observe $Y_{01}$ as responses from this arm (see Table~\ref{tab-strataDefns2}).
We label the $\phi$ parameters as before except that the arm index is now one of three: 'v', 'p', or 'i'.
Although we can observe $Y_{01}$, we do not observe $M_1$ in the arm with the infused mediator.
Thus, we can only estimate $\phi_{if}=\phi_{iaf} + \phi_{inf}$ and $\phi_{is}=\phi_{ias}+\phi_{ins}$.
Table~\ref{tab-strataDefns2} gives $\phi_{if}$ and
$\phi_{is}$ in terms of $\pi$ parameters, and using arguments similar to those used in Supplement Section~S3,
those extra $\phi$ parameters do not lead to identification of
$E \left[ Y_{1 M_0} \right]$ (see equation~\ref{eq:EY1M0.eq.EY1M1+e3+e4}) or
$E \left[ Y_{0 M_1} \right]$ (see equation~\ref{eq:EY0M1.eq.EY1M1+b+e2+e4}).

[Table 4 about here.]

%

Suppose that in addition to the base model assumptions, we can perfectly predict from baseline covariates which individuals will have $M_1=1$. Then
in the passive immunization arm
the expected proportion of failures among those predicted to have $M_1=1$ will be identified
as $Pr \left[ Y_{01}=1 \mbox{ and }  M_1=1 \right] = $ $\phi_{iaf}  = $ $  \pi_{0011}^{10} + \pi_{0111}^{10} + \phi_{vaf},$
while in the placebo arm the expected proportion of failure among those predicted to have $M_1=0$ will be identified as
$Pr \left[ Y_{00}=1 \mbox{ and }  M_1=0 \right]   = $ $\phi_{vnf} + \pi_{00 \cdot 1}^{00}.$
Thus,  we identify $E \left[ Y_{0M_1} \right]$ from the 3 arm trial,
$E \left[ Y_{0M_1} \right]  = $ $\sum_{m=0}^{1} Pr \left[ Y_{0m}=1 \mbox{ and }  M_1=m \right],$
 allowing us to identify $\theta_{I_a}$ and $\pra$. What makes this identification possible is that there are only 2 levels for $M_1$ and
 all individuals randomized to placebo get one level, while all those randomized to passive immunization get the other level. Identification with more than two levels
 for $M_1$ is not as straightforward.

\section{Nonbinary Mediators and adding Passive Immunity Experiments}
\label{sec-nonbinaryMediators}

In this section, we again allow a more general antibody response such that the support of $M_1$, $\mathcal{S}_{M_1}$, contains both undetectable (i.e., $M_1=0^*$)
or any of a  set of detectable antibody levels ($M_1 > 0^*$).
We continue to assume (similar to the base model of the binary case) that $Pr[ M_0=0^*]=1$.
We explore identification of $\pra$ from several different experimental designs, all of which allow three types of intervention, vaccine, placebo, and passive immunization.

\subsection{Three Arm Trial}
\label{sec-threearm.general}

Consider a randomized vaccine trial with three arms: placebo, vaccine, and passive immunization.
We denote the three arms as $p$, $v$, and $i$, with the  associated values of $A$ equal to $0$, $1$ and $2$,
and similarly for  antibody potential responses ($M_0$, $M_1$, and $M_2$) and potential outcomes ($Y_{0m},$ $Y_{1m}$, and $Y_{2m}$).

We start by reviewing some results
in \citet{Gilb:2022f}, who studied controlled vaccine efficacy from a placebo controlled randomized vaccine trial,
and also considered the case when  $Pr[ M_0=0^*]=1$.
\citet{Gilb:2022f} defined the controlled vaccine efficacy, which in our notation (i.e., when $Y_{am}$ is binary) is
$CVE(m_1,m_0)  = $ $1 - \frac{ E \left[ Y_{1m_1}  \right] }{ E \left[ Y_{0m_0}  \right] }.$
When $Pr[ M_0=0^*]=1$,  we write
\begin{eqnarray*}
CVE(m,0^*) \equiv CVE(m )  & = &  1 - \frac{ E \left[ Y_{1m}  \right] }{ E \left[ Y_{00}  \right] } \equiv  1 - \theta_{C}(m),
\end{eqnarray*}
where $Y_{00}$ is the placebo response when $m=0^*$. The $CVE(m)$ used $E(Y_{1m})$, the expected value of the potential response  $Y_{am}$ where $A$ is set to $a$ and $M$ is set of $m$.
To identify  $E(Y_{1m})$ from a random sample where individuals are vaccinated and each individual has their natural mediator value, we need to make some strong assumptions.
Let $X$ be baseline covariates such that the sequential ignorability assumptions, $SI_1$ and $SI_2$ hold, and assume the following positivity assumption holds,
{\it
\begin{description}
\item[$PosM1(  \mathcal{S}_{M_1})$:]
$Pr[ M_1=m | A=1, X=x] > 0 \mbox{ for all $x$ and } m \in \mathcal{S}_{M_1}.$
\end{description}
}
Then
\begin{eqnarray}
1 - \theta_{C}(m)= CVE(m )  & = &  1 - \frac{ \sum_{(m,x)} E \left[ Y_{1m} | X=x, M_1=m \right] Pr[ M_1=m \mbox{ and } X=x] }{ E \left[ Y_{00}  \right] },
\label{eq:CVE}
\end{eqnarray}
where the summation is over the support of $X$ and $M_1$ combined, and we assume discrete support for ease of exposition.
Then $\theta_{C}(m)$ is identifiable  for each $m \in \mathcal{S}_{M_1}$.
See \citet{Gilb:2022f} for details of identification results, inferential methods, and sensitivity analyses.

We can define an analogous controlled protective efficacy by comparing the passive immunization arm to placebo,
\begin{eqnarray}
CPE(m )  & = &  1 - \frac{ E \left[ Y_{2m}  \right] }{ E \left[ Y_{00}  \right] } \equiv  1 - \theta_{I_a}(m),
\label{eq:CPE}
\end{eqnarray}
where we use the notation $\theta_{I_a}(m)=1-CPE(m)$ because here
 $Y_{2m}$ (potential outcome from passive immunization where $M$ is set to $m$)
is acting like $Y_{0m}$ (potential outcome from placebo recipient where $M$ is set to $m$).
For the passive immunization arm, we can randomly assign individuals to $M_2$ with a known distribution
with support $\mathcal{S}_{M_1}$. In other words, we ensure by design that $Pr[ M_2=m|A=2]>0$ for each $m \in \mathcal{S}_{M_1}$
and $M_2 \ind Y_{am}$ instead of assuming sequential ignorability and positivity.

Since $\theta_C(m)$ and $\theta_{I_a}(m)$ are acting like a total effect and an adding indirect effect at $m$, we can define
the proportion of the controlled ratio effect at $m$ due to the controlled indirect ratio effect at $m$ as
\begin{eqnarray}
\pra(m) = \frac{ \log \left\{ \theta_{I_a}(m) \right\} }{ \log \left\{  \theta_C(m) \right\} }.
\label{eq:pram}
\end{eqnarray}
 These are useful estimands themselves, but they can also be used to identify $\theta_T$ and $\theta_{I_a}$.
Because the distribution of $M_1$ is identifiable from the vaccine arm, and because the denominator of the ratios in equations~\ref{eq:CVE} and \ref{eq:CPE} do not depend on $m$,
we get
\begin{eqnarray}
\theta_{\TE} = \sum_{m} \theta_{C}(m) Pr \left[ M_1 = m  \right] \label{eq:TE.sumCE}
\end{eqnarray}
and
\begin{eqnarray}
\theta_{\IE_a} = \frac{ E \left[ Y_{0M_1} \right] }{ E \left[ Y_{0 M_0} \right]} =  \sum_{m} \theta_{\IE_a}(m) Pr \left[ M_1 = m \right], \label{eq:TE.sumIEa}
\end{eqnarray}
and $\pra = \log( \theta_{I_a})/ \log(\theta_T)$, where the summations are over $\mathcal{S}_{M_1}$.

For this section, we have identified $\pra$ using the controlled vaccine efficacy and its analog for passive immunization, controlled protective efficacy. Since this is a randomized trail, both $SI_1$ and $PosA$
are met for both CVE and CPE, but  CVE additionally requires the $SI_2$
and $PosM1(\mathcal{S}_{M_1})$ as untestable assumptions, while the CPE can meet those later assumptions by design.

\subsection{Three Arm Trial with Closeout Vaccination}
\label{sec-threearm-X.general}

 \citet{Foll:2006} considered closeout vaccination, where individuals in the placebo arm are vaccinated at the end of the response follow-up period,
 to identify $M_1$ in the placebo arm without using baseline variables.
We modify that idea for the three arm trial.

To start, as in the previous section, we assume $E(Y_{0M_1}) = E(Y_{2M_1})$.
We  rewrite $E(Y_{2M_1})$ as
\begin{eqnarray}
E[Y_{2M_1}] & = & \sum_m       Pr(Y_{2m}=1 | M_1=m)  \times Pr(M_1=m)   \nonumber  \\
           & = & \sum_m \frac{Pr(M_1=m|Y_{2m}=1) Pr(Y_{2m}=1 )}{ Pr(M_1=m)} \times Pr(M_1=m) \nonumber  \\
           & = & \sum_m      Pr(M_1=m|Y_{2m}=1) Pr(Y_{2m}=1 )  \nonumber   \\
           & = & \sum_m  \left\{ Pr[ M_1=m] -     Pr(M_1=m|Y_{2m}=0) Pr(Y_{2m}=0 )  \right\}, \label{eq:EY0M1.Bayes}
\end{eqnarray}
where the second step uses Bayes theorem and the last step uses
\begin{eqnarray*}
P(M_1=m) & = & P(M_1=m|Y_{2m}=1) P(Y_{2m}=1) + P(M_1=m|Y_{2m}=0)  P(Y_{2m}=0).
\end{eqnarray*}

As in the previous section, we independently draw $M_2$ from a known distribution with support $\mathcal{S}_{M_1}$, such that $M_2$ is independent of all potential mediators and outcomes,
and    $Pr[ M_2=m | A=2]>0$ for each $m  \in \mathcal{S}_{M_1}$.
Then by  independence and consistency,
\begin{eqnarray*}
Pr[ Y_{2m}=0 ] & = &  Pr[ Y_{2m}=0 | M_2=m] = Pr[ Y | A=2, M_2=m],
\end{eqnarray*}
and $Pr[Y_{2m}=0]$ is identifiable. Also, $Pr[ M_1=m]$ is identifiable from the vaccine arm. The only remaining
piece to identify in equation~\ref{eq:EY0M1.Bayes} is $Pr(M_1=m|Y_{2m}=0)$, which we identify using closeout vaccination
in the passive immunization arm.

At the end of the response follow-up period, we implement a closeout vaccination on  individuals in the  passive immunization arm with $Y_{2M_2}=0$.
We do not vaccinate  individuals with  $Y_{2M_2}=1$  because the antibody
response after the getting the disease and vaccination cannot substitute for $M_1$.
We assume that on the closeout vaccinated individuals the antibody response after closeout vaccination would equal $M_1$, the value they would have gotten if vaccinated at the
start of the trial.
This may be a tenuous assumption if the response is disease and there are asymptomatic infections that have  $Y_{2M_2}=0$, since
asymptomatic infections may affect a subsequent immune response to vaccination.
We do not assume that $M_2=M_1$. Instead, we use the independence and positivity assumptions on $M_2$ previously mentioned to get
\begin{eqnarray*}
Pr[ M_1=m | Y_{2m}=0 ] & = & Pr[ M_1=m | Y_{2m}=0, M_2=M_1 ],
\end{eqnarray*}
so that $Pr[ M_1=m | Y_{2m}=0 ]$ is identifiable. Thus, by equation~\ref{eq:EY0M1.Bayes} we can identify $E[Y_{2M_1}]$,
and hence also identify $\theta_{I_a}$.

In summary, a three arm trial with closeout vaccination of the passive immunization arm
allows identification of $\theta_{I_a}$ and $\lambda_a$ without the use of any baseline
covariates, without assuming $SI_2$, and allowing $Pr[ M_0=0^*]=1$.

\subsection{Two Randomized Trials}
\label{sec-controlled}

In this section, we consider combining two randomized trials, a vaccine vs. placebo (VP)   trial  and
a passive immunization vs.  placebo (IP) trial. This requires more care because there will be differences between the trials,
such as distributions for exposure, baseline covariates, potential mediator responses,  and potential outcomes.
To emphasize these differences, we use superscripts VP and IP to differentiate the two trials in the variables.

Suppose that using the VP trial, we can identify baseline predictors of $M_1$, such that for each $m \in \mathcal{S}_{M_1}$
there is a predictor set of baseline variables,  $\mathcal{S}_x(m) = \left\{ x: \mbox{ if $X=x$ then } M_1=m \right\}$.
Further, assume that $\mathcal{S}_x(m)$ can identify participants in the IP trial with $M_1=m$.
Then we design the IP trial using a quota sampling approach so that we chose
the participants for the IP trial
from the pool of available volunteers such that the baseline distribution of $X$ is similar between the two trials.
For all participants in the IP trial with $X \in S_{x}(m)$, we set $M_2=m$.
If the sampling is done well enough,
then the distribution of $M_1$ will match the distribution of $M_2$ and
\begin{eqnarray}
\theta_{I_a} & = & \frac{ E \left[ Y_{0M_1}^{VP} \right] }{E \left[ Y_{0M_0}^{VP} \right]} \approx \frac{ E \left[ Y_{2M_2}^{IP} \right] }{E \left[ Y_{0M_0}^{IP} \right]}. \label{eq:thetaIa.2trials.Approach1}
\end{eqnarray}
In expression~\ref{eq:thetaIa.2trials.Approach1} the approximation may be reasonable because even though we will have different exposure distributions in the two trials,
the exposure effects should cancel out  as in equation~\ref{eq:VE.ZYstar}.

A second approach is
similar to that of Section~\ref{sec-threearm.general}.
Rewrite equation~\ref{eq:CVE} from the VP trial using the new notation,
\begin{eqnarray}
CVE(m)  & = &  1 - \frac{ \sum_{x} E \left[ Y_{1m}^{VP} | X^{VP}=x  \right] Pr[ X^{VP}=x ] }{ E \left[ Y_{00}^{VP} \right] } = 1-\theta_C(m),
\label{eq:CVE2}
\end{eqnarray}
and make the same sequential ignorability and positivity assumptions as in Section~\ref{sec-threearm.general}.
For the IP trial, we assume that within levels of $X$, the study populations between the trials are comparable and capture all the differences between the trials
except the exposure effects, which we assume cancel out by equation~\ref{eq:VE.ZYstar}. Then
we modify equation~\ref{eq:CPE} to standardize using the distribution of $X$ from the VP trial,
\begin{eqnarray}
CPE(m) & = & 1 - \frac{ \sum_{x} E \left[ Y_{2m}^{IP} | X^{IP}=x  \right] Pr[ X^{VP}=x ] }{ E \left[ Y_{00}^{IP} | X^{IP} \right] Pr[ X^{VP}=x ] } = 1- \theta_{I_a}(m),
\label{eq:CPE2}
\end{eqnarray}
where now we must standardize in both the numerator and denominator.  As in Section~\ref{sec-threearm.general}, we can impose the independence of the $m$ and potential outcomes in the $A=2$ arm
by randomization in the design, not by assumption.  In summary, we are controlling for differences in $X$ between the trials by standardization,
and controlling for differences in exposure by equation~\ref{eq:VE.ZYstar}.
We again define $\pra(m)$ as in equation~\ref{eq:pram}, and because the denominators still do not depend on $m$,
we can get $\theta_{\TE}$ and $\theta_{I_a}$ by equations~\ref{eq:TE.sumCE} and \ref{eq:TE.sumIEa} respectively.



\begin{figure}

\includegraphics[width=7in]{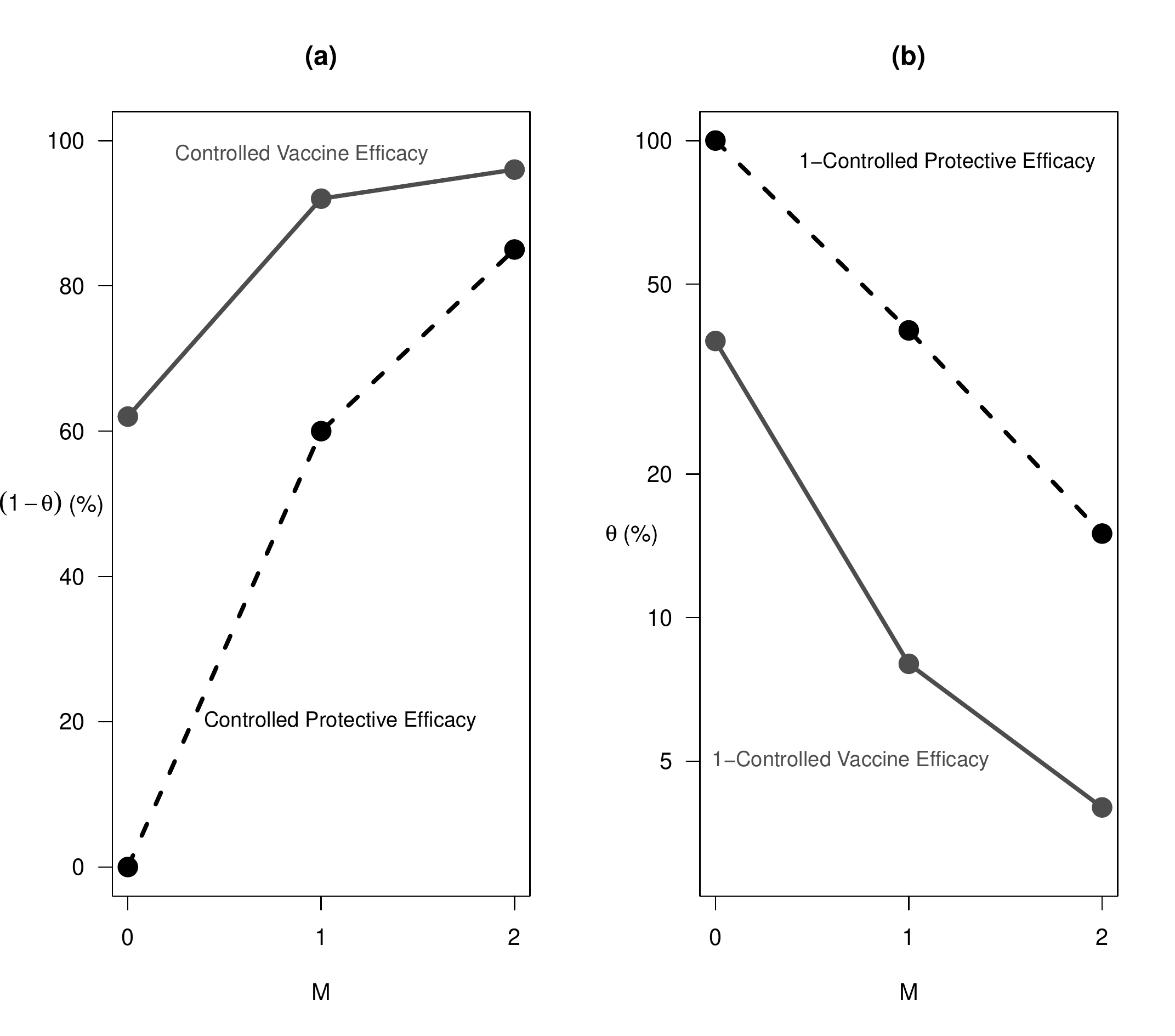}

\caption{
Example for combining two trials, a vaccine vs. placebo (VP) trial and a passive immunization vs. placebo (IP) trial.
In panel (a) we plot the controlled vaccine efficacy ($CVE(m)=1 - \theta_C(m)$) for the VP trial and the controlled protective efficacy ($CPE(m)=1-\theta_{I_a}(m)$) for the IP trial,
while in panel (b) we plot the same data in terms of ratio effects ($\theta_C(m)$ for the VP trial, and $\theta_{I_a}(m)$ for the IP trial) but on a log scale.
The efficacies (i.e., vertical axes) are in percent.
\label{fig-2trials}}

\end{figure}

Consider making inferences using equations~\ref{eq:CVE2} and \ref{eq:CPE2}.
In Table~\ref{tab-2trials} and Figure~\ref{fig-2trials} we give a made-up example assuming the covariate adjustments of equations~\ref{eq:CVE2} and \ref{eq:CPE2} have already been made.
In this example, there are 3 levels of the Ab mediator ($m=0,1,2$). At $m=0$ the value of $\hat{\theta}_{I_a}(0)=100\%$ so that $\hat{\lambda}_a(0)=0$.
The overall effects are estimated by weighted averages, so that using equation~\ref{eq:TE.sumCE} we get $\hat{\theta}_T$ and using equation~\ref{eq:TE.sumIEa} we get $\hat{\theta}_{I_a}$.
The overall $\pra$ is then estimated by $\hat{\lambda}_a = \log( \hat{\theta}_{I_a} )/ \log( \hat{\theta}_T )$. Compare  $\hat{\theta}_{I_a}=45.4\%$ and $\hat{\lambda}_a=45.4\%$
from Table~\ref{tab-2trials}
to  $\hat{\theta}_{I_s} = 0.09/0.38 = 23.7\%$  and $\hat{\lambda}_s=\log(0.38)/\log(0.09)=40.2\%$ calculated using assumption $SI_2$ and equation~\ref{eq:EY1M0}
which gives $\hat{E} \left[ Y_{1M_0} \right] = 38\%$. So if $SI_2$ is true and the sample size is large enough that we can ignore variability of the estimators, then $\theta_{I_s} < \theta_{I_a}$ and there is an interaction (since $\theta_{I_s} \neq \theta_{I_a}$).

[Figure 2 and Table 5 about here]

\section{Discussion}

We have explored the assumptions related to identifying the proportion of the total ratio effect  mediated through the antibodies
in a vaccine versus placebo (VP) trial. We have shown that because the placebo arm will likely not have positive antibody values, from the VP trial we can only identify
one of the two ways to define that proportion ($\prs$), and that requires an independence assumption between potential mediators and potential outcomes.
We discussed supplementing the VP trial with extra experimental information: either adding to the VP trial a third arm with a passive immunization, or combining the VP trial with
a passive immunization versus placebo (IP) trial.
These newly proposed experimentally supplemented trials allow us to identify the proportion of the total ratio effect due to adding antibodies to the placebo arm, $\pra$.

We note several differences between the traditional (i.e., VP trial alone) and non-traditional (i.e., VP trial with supplemental experimental passive immunization) mediation analyses.
First, the traditional mediation analysis is estimating $\prs$, while the non-traditional one is estimating $\pra$.
If we can make the simplifying assumption that the size of the effect of adding antibodies to participants in the placebo arm is the same
as that of subtracting antibodies from participants in the vaccine arm, then $\pr= \prs=\pra$, and the proposed non-traditional supplemental
experiments provide a way to estimate $\pr$ in a different way. For the passive immunization arm, we do not need to assume the
antibodies are independent of the potential outcomes because we can impose that independence by actively randomizing participants to their antibody values.
 In practice, we can do both mediation analyses and can compare estimates of both $\prs$ and $\pra$.
 If they are similar, then it may be reasonable to assume no interaction such that $\prs=\pra$ and they are two estimators of the same parameter, $\pr$.

Even if we cannot assume $\prs=\pra$, an advantage of estimating $\pra$ using the non-traditional mediation analyses is that the cross-world parameter $E(Y_{0M_1})$
is estimated experimentally (see e.g., Section~\ref{sec-threearm-X.general}). In contrast, the cross-world parameter
$E(Y_{1M_0})$ used in traditional mediation analyses for estimating $\prs$ requires making necessary and often difficult independence assumptions, which ideally
should be accompanied with sensitivity analyses (see similar sensitivity analyses for controlled vaccine efficacy in \cite{Gilb:2022f}).
In practice, some thought is required to accept the consistency assumptions  with respect to the vaccine-induced antibodies having an equal effect as monoclonal antibodies.
It could be the
the monoclonal antibodies are made from a different virus strain that the vaccine-induced ones, and the monoclonal antibodies may have different immunological characteristics.
Further, the decay
pharmacokinetics are likely to be different between the two types of antibodies so timing of antibody measurements is important for these types of studies.

In this paper, we have focused on  vaccine studies where very few or no placebo participants would produce antibodies.
We have also focused almost exclusively on examining assumptions and identifiability. We explored different study designs
to identify these mediation effects. One design in particular
(a three arm randomized trial with closeout vaccination, see Section~\ref{sec-threearm-X.general})
is feasible and does not require as severe assumptions as the others.
There is much room for future work in spelling out the details of that design and the other proposed
 study designs, including examining how the designs may need to be modified for feasibility reasons.
Future work could explore these issues when the some placebo
participants have existing antibodies due to natural exposure.
Other work is needed to explore
 details of estimators (e.g., \cite{Benk:2021}), and this paper could be complementary to some of that work,
since the estimator work typically does not have the space to fully discuss the implications of all their assumptions.

\section{Supplementary Material}

See Supplementary Material for proofs and extra mathematical details.

\section*{Acknowledgements}

We thank Peter Gilbert and three anonymous reviewers for insightful comments that appreciably improved the article.

\bibliographystyle{unsrtnat}

\bibliography{mediationrefs}

\begin{thebibliography}{23}
\providecommand{\natexlab}[1]{#1}
\providecommand{\url}[1]{\texttt{#1}}
\expandafter\ifx\csname urlstyle\endcsname\relax
  \providecommand{\doi}[1]{doi: #1}\else
  \providecommand{\doi}{doi: \begingroup \urlstyle{rm}\Url}\fi

\bibitem[Halloran et~al.(2010)Halloran, Longini, Struchiner, and
  Longini]{Hall:2010book}
M~Elizabeth Halloran, Ira~M Longini, Claudio~J Struchiner, and Ira~M Longini.
\newblock \emph{Design and analysis of vaccine studies}, volume~18.
\newblock Springer, 2010.

\bibitem[VanderWeele(2015)]{Vand:2015}
Tyler VanderWeele.
\newblock \emph{Explanation in causal inference: methods for mediation and
  interaction}.
\newblock Oxford University Press, 2015.

\bibitem[Gilbert et~al.(2022{\natexlab{a}})Gilbert, Montefiori, McDermott,
  Fong, Benkeser, Deng, Zhou, Houchens, Martins, Jayashankar,
  et~al.]{Gilb:2022}
Peter~B Gilbert, David~C Montefiori, Adrian~B McDermott, Youyi Fong, David
  Benkeser, Weiping Deng, Honghong Zhou, Christopher~R Houchens, Karen Martins,
  Lakshmi Jayashankar, et~al.
\newblock Immune correlates analysis of the mrna-1273 covid-19 vaccine efficacy
  clinical trial.
\newblock \emph{Science}, 375:\penalty0 43--50, 2022{\natexlab{a}}.

\bibitem[Benkeser et~al.(2021)Benkeser, Diaz, and Ran]{Benk:2021}
D.~Benkeser, I.~Diaz, and J.~Ran.
\newblock Inference for natural mediation effects under case-cohort sampling
  with applications in identifying covid-19 vaccine correlates of protection.
\newblock \emph{arXiv:2103.02643v1}, 2021.

\bibitem[Robins and Greenland(1992)]{Robi:1992}
James~M Robins and Sander Greenland.
\newblock Identifiability and exchangeability for direct and indirect effects.
\newblock \emph{Epidemiology}, pages 143--155, 1992.

\bibitem[Seigrist and Lambert(2016)]{Seig:2016}
C.A. Seigrist and P.H. Lambert.
\newblock Chapter 2: How vaccines work.
\newblock In BR~Bloom and PH~Lambert, editors, \emph{The Vaccine Book}, pages
  33--42. Elsevier, New York, 2016.

\bibitem[Rapaka et~al.(2022)Rapaka, Hammershaimb, and Neuzil]{Rapa:2022}
Rekha~R Rapaka, Elizabeth~A Hammershaimb, and Kathleen~M Neuzil.
\newblock Are some covid-19 vaccines better than others? interpreting and
  comparing estimates of efficacy in vaccine trials.
\newblock \emph{Clinical Infectious Diseases}, 74\penalty0 (2):\penalty0
  352--358, 2022.

\bibitem[Cowling et~al.(2019)Cowling, Lim, Perera, Fang, Leung, Peiris, and
  Tchetgen~Tchetgen]{Cowl:2019}
Benjamin~J Cowling, Wey~Wen Lim, Ranawaka~APM Perera, Vicky~J Fang, Gabriel~M
  Leung, JS~Malik Peiris, and Eric~J Tchetgen~Tchetgen.
\newblock Influenza hemagglutination-inhibition antibody titer as a mediator of
  vaccine-induced protection for influenza b.
\newblock \emph{Clinical Infectious Diseases}, 68\penalty0 (10):\penalty0
  1713--1717, 2019.

\bibitem[Nguyen et~al.(2015)Nguyen, Osypuk, Schmidt, Glymour, and
  Tchetgen~Tchetgen]{Nguy:2015}
Quynh~C Nguyen, Theresa~L Osypuk, Nicole~M Schmidt, M~Maria Glymour, and Eric~J
  Tchetgen~Tchetgen.
\newblock Practical guidance for conducting mediation analysis with multiple
  mediators using inverse odds ratio weighting.
\newblock \emph{American journal of epidemiology}, 181\penalty0 (5):\penalty0
  349--356, 2015.

\bibitem[Hudgens et~al.(2004)Hudgens, Gilbert, and Self]{Hudg:2004}
Michael~G Hudgens, Peter~B Gilbert, and Steven~G Self.
\newblock Endpoints in vaccine trials.
\newblock \emph{Statistical methods in medical research}, 13\penalty0
  (2):\penalty0 89--114, 2004.

\bibitem[Senn(2022)]{Senn:2022}
Stephen Senn.
\newblock The design and analysis of vaccine trials for covid-19 for the
  purpose of estimating efficacy.
\newblock \emph{Pharmaceutical Statistics}, 21\penalty0 (4):\penalty0 790--807,
  2022.

\bibitem[Cole and Frangakis(2009)]{Cole:2009}
Stephen~R Cole and Constantine~E Frangakis.
\newblock The consistency statement in causal inference: a definition or an
  assumption?
\newblock \emph{Epidemiology}, 20\penalty0 (1):\penalty0 3--5, 2009.

\bibitem[Hafeman and Schwartz(2009)]{Hafe:2009}
Danella~M Hafeman and Sharon Schwartz.
\newblock Opening the black box: a motivation for the assessment of mediation.
\newblock \emph{International journal of epidemiology}, 38\penalty0
  (3):\penalty0 838--845, 2009.

\bibitem[Pearl(2001)]{Pear:2001}
Judea Pearl.
\newblock Direct and indirect effects.
\newblock In \emph{Proceedings of the Seventeenth Conference on Uncertainty in
  Artificial Intelligence}, pages 411--420. Morgan Kaufmann, San Francisco,
  2001.

\bibitem[Judea(2012)]{Pear:2012}
Pearl Judea.
\newblock The mediation formula: A guide to the assessment of causal pathways
  in nonlinear models.
\newblock In C.~Berzuini, P.~Dawid, and L.~Bernardinelli, editors,
  \emph{Causality: Statistical Perspectives and Applications}, pages 151--179.
  Wiley, 2012.

\bibitem[Imai et~al.(2010)Imai, Keele, and Yamamoto]{Imai:2010}
Kosuke Imai, Luke Keele, and Teppei Yamamoto.
\newblock Identification, inference and sensitivity analysis for causal
  mediation effects.
\newblock \emph{Statistical science}, 25\penalty0 (1):\penalty0 51--71, 2010.

\bibitem[Goel et~al.(2021)Goel, Apostolidis, Painter, Mathew, Pattekar,
  Kuthuru, Gouma, Hicks, Meng, Rosenfeld, et~al.]{Goel:2021}
Rishi~R Goel, Sokratis~A Apostolidis, Mark~M Painter, Divij Mathew, Ajinkya
  Pattekar, Oliva Kuthuru, Sigrid Gouma, Philip Hicks, Wenzhao Meng, Aaron~M
  Rosenfeld, et~al.
\newblock Distinct antibody and memory b cell responses in sars-cov-2
  na{\"\i}ve and recovered individuals after mrna vaccination.
\newblock \emph{Science immunology}, 6\penalty0 (58):\penalty0 eabi6950, 2021.

\bibitem[Hafeman and VanderWeele(2011)]{Hafe:2011}
Danella~M Hafeman and Tyler~J VanderWeele.
\newblock Alternative assumptions for the identification of direct and indirect
  effects.
\newblock \emph{Epidemiology}, pages 753--764, 2011.

\bibitem[Shukla et~al.(2020)Shukla, Ramasamy, Shanmugam, Ahuja, and
  Khanna]{Shuk:2020}
Rahul Shukla, Viswanathan Ramasamy, Rajgokul~K Shanmugam, Richa Ahuja, and
  Navin Khanna.
\newblock Antibody-dependent enhancement: a challenge for developing a safe
  dengue vaccine.
\newblock \emph{Frontiers in Cellular and Infection Microbiology}, 10:\penalty0
  Article 572681, 2020.

\bibitem[Gilbert et~al.(2022{\natexlab{b}})Gilbert, Fong, Kenny, and
  Carone]{Gilb:2022f}
Peter~B Gilbert, Youyi Fong, Avi Kenny, and Marco Carone.
\newblock A controlled effects approach to assessing immune correlates of
  protection.
\newblock \emph{Biostatistics}, DOI:\penalty0 10.1093/biostatistics/kxac24,
  2022{\natexlab{b}}.

\bibitem[Follmann(2006)]{Foll:2006}
Dean Follmann.
\newblock Augmented designs to assess immune response in vaccine trials.
\newblock \emph{Biometrics}, 62\penalty0 (4):\penalty0 1161--1169, 2006.

\bibitem[Naimi et~al.(2014)Naimi, Kaufman, and MacLehose]{Naim:2014}
Ashley~I Naimi, Jay~S Kaufman, and Richard~F MacLehose.
\newblock Mediation misgivings: ambiguous clinical and public health
  interpretations of natural direct and indirect effects.
\newblock \emph{International journal of epidemiology}, 43\penalty0
  (5):\penalty0 1656--1661, 2014.

\bibitem[VanderWeele(2016)]{Vand:2016}
Tyler~J VanderWeele.
\newblock Mediation analysis: a practitioner's guide.
\newblock \emph{Annual review of public health}, 37:\penalty0 17--32, 2016.

\end{thebibliography}

\begin{table}

\caption{
The 12 types left after excluding types based on assumptions $RG_1,RG_2$ and $RG_3$, and those with $M_0=1$.
Values in gray boxes are not observable in the randomized trial (i.e., are cross-world counterfactuals).
The notation for the proportion of the population for each type  is $\pi^{\bf M}_{\bf Y}$, where ${\bf M}= [M_1, M_0]$ and ${\bf Y} = [Y_{11}, Y_{10}, Y_{01}, Y_{00}]$.
In the description, the ``either'' or ''both'' refer to the antibody (Ab)
and/or the direct effect,  prot.=protected and resp.=responder.
\label{tab-MY}
}

\begin{center}
\begin{tabular}{|l|cccccc||rrrr|ll|} \hline
 $\pi_{Y_{11}  Y_{10}  Y_{01}  Y_{00} }^{M_1  M_0}$ & $M_1$ & $M_0$ & $Y_{11}$ & $Y_{10}$ & $Y_{01}$ & $Y_{00}$  & $Y_{1M_1}$ & $Y_{1M_0}$ & $Y_{0M_1}$  & $Y_{0M_0}$  & \multicolumn{2}{c|}{Type Description} \\ \hline
$\pi_{0000}^{00}$ & 0 & 0 & \colorbox{lightgray}{0} & 0 & \colorbox{lightgray}{0} & 0   & 0 & 0 & 0 & 0 & Ab nonresp. & uninfectable\\
$\pi_{0001}^{00}$ & 0 & 0 & \colorbox{lightgray}{0} & 0 & \colorbox{lightgray}{0} & 1     & 0 & 0 & 1 & 1 & Ab nonresp. & either alone prot.  \\
$\pi_{0011}^{00}$ & 0 & 0 & \colorbox{lightgray}{0} & 0 & \colorbox{lightgray}{1} & 1    & 0 & 0 & 1 & 1 & Ab nonresp. & direct alone prot. \\
$\pi_{0101}^{00}$ & 0 & 0 & \colorbox{lightgray}{0} & 1 & \colorbox{lightgray}{0} & 1    & 1 & 1 & 1 & 1 & Ab nonresp. & Ab alone prot. \\
$\pi_{0111}^{00}$ & 0 & 0 & \colorbox{lightgray}{0} & 1 & \colorbox{lightgray}{1} & 1    & 1 & 1 & 1 & 1 & Ab nonresp. & need both
\\
$\pi_{1111}^{00}$ & 0 & 0 & \colorbox{lightgray}{1} & 1 & \colorbox{lightgray}{1} & 1   & 1 & 1 & 1 & 1 & Ab nonresp. & totally doomed
\\
$\pi_{0000}^{10}$ & 1 & 0 & 0 & \colorbox{lightgray}{0} & \colorbox{lightgray}{0} & 0   & 0 & \colorbox{lightgray}{0} & \colorbox{lightgray}{0} & 0 & Ab resp. & uninfectable\\
$\pi_{0001}^{10}$ & 1 & 0 & 0 & \colorbox{lightgray}{0} & \colorbox{lightgray}{0} & 1   & 0 & \colorbox{lightgray}{0} & \colorbox{lightgray}{0} & 1 & Ab resp. & either alone prot. \\
$\pi_{0011}^{10}$ & 1 & 0 & 0 & \colorbox{lightgray}{0} & \colorbox{lightgray}{1} & 1   & 0 & \colorbox{lightgray}{0} & \colorbox{lightgray}{1} & 1 & Ab resp. & direct alone prot. \\
$\pi_{0101}^{10}$ & 1 & 0 & 0 & \colorbox{lightgray}{1} & \colorbox{lightgray}{0} & 1   & 0 & \colorbox{lightgray}{1} & \colorbox{lightgray}{0} & 1 & Ab resp. & Ab alone prot. \\
$\pi_{0111}^{10}$ & 1 & 0 & 0 & \colorbox{lightgray}{1} & \colorbox{lightgray}{1} & 1   & 0 & \colorbox{lightgray}{1} & \colorbox{lightgray}{1} & 1 & Ab resp. & need both  \\
$\pi_{1111}^{10}$ & 1 & 0 & 1 & \colorbox{lightgray}{1} & \colorbox{lightgray}{1} & 1   & 1 & \colorbox{lightgray}{1} & \colorbox{lightgray}{1} & 1 & Ab resp. & totally doomed
\\ \hline
\end{tabular}
\end{center}

\end{table}

\begin{table}

\caption{
The proportions of the twelve types of responses in the base model, and the marginal  combinations of them.
Let ${\bf M}=[M_1, M_0]$ and ${\bf Y}=[Y_{11}, Y_{10}, Y_{01}, Y_{00}]$. The proportions are written as
$\pi_{\bf Y}^{\bf M}$. For example, $\pi^{10}_{0101}$ is the proportion of the population with potential mediators equal to $M_1=1$ and $M_0=0$, and
potential outcomes equal to
 $Y_{11}=0$, $Y_{10}=1$, $Y_{01}=0$ and $Y_{00}=1$.  We only observe six marginals from the table (4 from the treatment arm [$A=1$], and 2 from the control arm [$A=0$]), so those combinations of types are identifiable, as well as certain functions of them (e.g., the sum of the marginals for the first two rows is $Pr[M_1=0]$, which is identifiable).
The observable marginals as written as $\phi_{amy}$, where the $_{amy}$ subscripts refer to respectively, $A$ ($a = v$ is {\bf v}accine arm, $a=p$ is {\bf p}lacebo arm),
$M_A$ ($m=a$ is detectable {\bf a}ntibodies, $m=n$ is {\bf n}o detectable antibodies), and $Y_{AM_A}$ ($y=f$ is {\bf f}ailure, $y=s$ is {\bf s}uccess).
\label{tab-strataDefns}
}

\begin{center}
\begin{tabular}{|l|c|ccccc|r|}  \hline
& $A=0, M_0=0,$  & \multicolumn{5}{c|}{$A=0, M_0=0, Y_{00}=1$   } &   \\
& $Y_{00}=0$  & \multicolumn{5}{c|}{ } &  observable \\
& ${\bf [0000]}$  & ${\bf [0001]}$ & ${\bf [0011]}$ & ${\bf [0101]}$ & ${\bf [0111]}$ & ${\bf [1111]}$ &  marginal \\ \hline
$A=1,M_1=0,Y_{10}=0$ &  $\pi_{0000}^{00}$ &  $\pi_{0001}^{00}$ & $\pi_{0011}^{00}$ &  &  &   &  $\phi_{vns}$ \\
$A=1,M_1=0,Y_{10}=1$ &   &  &  &  $\pi_{0101}^{00}$ & $\pi_{0111}^{00}$ & $\pi_{1111}^{00}$  &  $\phi_{vnf}$ \\
$A=1,M_1=1,Y_{11}=0$ &  $\pi_{0000}^{10}$ &  $\pi_{0001}^{10}$ & $\pi_{0011}^{10}$ &  $\pi_{0101}^{10}$ & $\pi_{0111}^{10}$ &   &  $\phi_{vas}$ \\
$A=1,M_1=1,Y_{11}=1$ &   &  &  &   &    & $\pi_{1111}^{10}$  &  $\phi_{vaf}$ \\ \hline
observable marginal &   $\phi_{pns}$ &    &   &    $\phi_{pnf}$ &   &    &    1  \\ \hline
\end{tabular}
\end{center}

\end{table}

\begin{table}
\caption{ Made-up results from a randomized  placebo-controlled vaccine trial with 10,000 participants in each arm.
\label{tab:toy}}

\begin{center}
\begin{tabular}{llllll}
   &   &    & Number of & $\phi$ & Conditional Event \\
$a$ & $m(a)$ & $y(a,m(a))$ & Participants & parameter & Proportions \\ \hline
1 & 1 & 1 & 2  & $\hat{\phi}_{vaf}=0.02\%$ &  $\hat{E}[Y_{ 1M_1} | M_1=1 ]= 2/8000 =0.025\%$ \\
1 & 1 & 0 & 7998 & $\hat{\phi}_{vas}=79.98\%$ & \\
1 & 0 & 1 & 8    & $\hat{\phi}_{vnf}=0.08\%$ & $\hat{E}[Y_{ 1M_1} | M_1=0]= 8/2000=0.4\%$ \\
1 & 0 & 0 & 1992 & $\hat{\phi}_{vns}=19.92\%$ & \\
0 & 1 & 1 & 0 & $\hat{\phi}_{paf}=0$ &  \\
0 & 1 & 0 & 0 & $\hat{\phi}_{pas}=0$ & \\
0 & 0 & 1 & 100 & $\hat{\phi}_{pnf}=1\%$ & $\hat{E}[Y_{0M_0} | M_0=0]= 100/10000=1\%$ \\
0 & 0 & 0 & 9900 & $\hat{\phi}_{pns}=99\%$ \\ \hline
\end{tabular}
\end{center}

\end{table}

\begin{table}

\caption{
Ten strata (4 strata in the vaccine arm,  4 in the placebo arm, and 2 in the passive immunization arm), where the 5th and 6th rows represent strata that are not observed when $M_0=0$ always (as in the base model).
The last 6 columns represent the observed values $(A,M_A,Y_{AM_A})$ put in the appropriate column, with `\cdt'  representing unobserved potential outcomes.
\label{tab-strataDefns2}
}

\begin{center}
\begin{tabular}{|lll|ccc|cccccc|}  \hline
description & $\phi$ & $\pi$  & $A$ & $M_A$ & $Y_{AM_A}$ & $M_1$ & $M_0$ & $Y_{11}$ & $Y_{10}$ & $Y_{01}$ & $Y_{00}$ \\  \hline
vaccine/Abs/failure & $\phi_{vaf}$ & $\pi_{1111}^{10}$ &  1 & 1  & 1           & 1   & \cdt &  1      & \cdt & \cdt & \cdt \\
vaccine/Abs/success & $\phi_{vas}$ & $\pi_{0000}^{10}+\pi_{0\cdot \cdot 1}^{10}$ & 1 & 1    & 0         &  1   & \cdt & 0      & \cdt & \cdt & \cdt \\
vaccine/no Abs/failure & $\phi_{vnf}$ & $\pi_{\cdot 1 \cdot 1}^{00}$ & 1 & 0  & 1           & 0   & \cdt & \cdt & 1  & \cdt  & \cdt  \\
vaccine/no Abs/success & $\phi_{vns}$ & $\pi_{0000}^{00}+\pi_{00\cdot 1}^{00}$ & 1 & 0  & 0           & 0   & \cdt & \cdt & 0  & \cdt  & \cdt  \\
placebo/Abs/failure & $\phi_{paf}$ & $0$ &  0 & 1  & 1           & \cdt   & 1 &   \cdt     & \cdt & 1 & \cdt \\
placebo/Abs/success & $\phi_{pas}$ & $0$ & 0 & 1    & 0         &  \cdt   & 1 &  \cdt      & \cdt & 0 & \cdt \\
placebo/no Abs/failure & $\phi_{pnf}$ & $\pi_{00\cdot 1}^{00}+\pi_{\cdot 1 \cdot 1}^{00}+\pi_{0\cdot \cdot 1}^{10}+\pi_{1111}^{10}$ & 0 & 0  & 1           & \cdt   & 0 & \cdt   & \cdt  & \cdt & 1   \\
placebo/no Abs/success & $\phi_{pns}$ & $\pi_{0000}^{00}+\pi_{0000}^{10}$ &  0 & 0  & 0           & \cdt   & 0 & \cdt   & \cdt  & \cdt &  0  \\
passive Imm/failure & $\phi_{if}$ & $\pi_{0011}^{00} + \pi_{0111}^{00} + \pi_{1111}^{00} +$  &  2  & \cdt & 1 & \cdt   & \cdt & \cdt   & \cdt  & 1 &  \cdt  \\
 &  &  $\pi_{0011}^{10} + \pi_{0111}^{10} + \pi_{1111}^{10}$ &   &  &  &  &  &    &  &  &    \\
passive Imm/success & $\phi_{is}$ & $\pi_{0000}^{00} + \pi_{0001}^{00} + \pi_{0101}^{00} + $ & 2 & \cdt & 0 & \cdt   & \cdt & \cdt   & \cdt  & 0 &  \cdt  \\
 &  & $\pi_{0000}^{10} + \pi_{0001}^{10} + \pi_{0101}^{10}$ &  &  &  &    &  &    &  &  &   \\
 \hline
\end{tabular}
\end{center}

\end{table}

\begin{table}[ht]
\centering

\caption{
Example for combining two trials, a vaccine vs. placebo (VP) trial and a passive immunization vs. placebo (IP) trial.
All values except Ab level are in percent.
\label{tab-2trials}
}

\begin{tabular}{rrrrrrr}
  \hline
  m (Ab level) & CVE(m) & CPE(m) & $\theta_C(m)$ & $\theta_{I_a}(m)$ & $\pra(m)$ & $P(M_1=m)$ \\
  \hline
   0 & 62.0 & 0.0 & 38.0 & 100.0 & 0.0 & 10 \\
   1 & 92.0 & 60.0 & 8.0 & 40.0 & 36.3 & 40 \\
   2 & 96.0 & 85.0 & 4.0 & 15.0 & 58.9 & 50 \\ \hline
Overall & 91.0 & 66.5 & 9.0 & 33.5 & 45.4 & 100
\end{tabular}
\end{table}

\clearpage

\Large{ \bf Supplementary Materials for: \\
Mediation Analyses for the Effect of Antibodies in  Vaccination
}

\large{Michael P. Fay and Dean A. Follmann}

\setcounter{page}{1}
\setcounter{section}{0}
\renewcommand{\thesection}{S\arabic{section}}
\setcounter{equation}{0}
\renewcommand{\theequation}{S\arabic{equation}}



\section{Pearl Assumptions}
\label{sec-PearlAssumptions}

We restate standard assumptions introduced by \citet{Pear:2001} used to identify direct or indirect effects, replacing ``exposure'' with ``vaccine'' for our example.
These 4 assumptions are listed in many publications; we choose the wording from  \citep{Naim:2014} for the first 3, and from  \citep{Vand:2016} for the last one.

\vspace*{2em}
\begin{tabular}{lll}
 & Text  & Mathematical  \\
 &   Description                & Notation  \\ \hline
 A1 &  ``No uncontrolled [vaccine]-outcome confounding'' &  $Y_{am} \ind A | C$ \\
 A2 &  ``No uncontrolled mediator-outcome confounding''  &  $Y_{am} \ind M_A | A,C$ \\
 A3 &  ``No uncontrolled [vaccine]-mediator confounding'' & $M_a \ind A | C$ \\
A4 &  ``No mediator-outcome confounding that is itself affected by the [vaccine]'' & $Y_{am} \ind M_{a'} | C$ \\
\end{tabular}
\vspace*{2em}

In the mathematical notation, the lack of a subscript denotes  that the independence expression holds for all individuals. The value $C$ denotes a vector of confounding variables (including possibly variables measured after baseline [i.e., after vaccination]).   Lower case letters indicate that the independence assumption holds if the values were set to a specific allowable constant.
For example,  $Y_{am} \ind A | C$ means that the potential outcome for individual $i$ given $A_i=a$ and their mediator was set to  $m$  (i.e., $Y^{(i)}_{am}$) is independent of $A_i$ given $C_i$ for all $a \in \left\{ 0,1 \right\}$, $m \in [0,\infty)$, and $i \in \left\{ 1, \ldots, n \right\}$.
Typically, the positivity assumptions are implicitly assumed.

\section{Proof of Equation~4
When ${ \bf \mathrm{Pr}[M_0 > 0^* | X=a, X=x]=0}$}
\label{app-EYaMb-Proof}

We loosely follow the proof of \citet[][p. 465]{Vand:2015}, except we focus on the binary mediator and outcome case with $a=1$ and $a'=0$ and we do not require that $Pr[ M_0>0^* | A=0,X]>0$. In the following,  we generalize the  $PosM_0$ and $PosM_1$
assumptions  as
\begin{description}
\item[$PosM_0(\mathcal{S}_{M_0|x})$:] $Pr \left[ M_0=m| A=0, X=x \right]>0$ for all $x$ and $m \in \mathcal{S}_{M_0|x}$, and
\item[$PosM_1(\mathcal{S}_{M_1|x})$:] $Pr \left[ M_1=m| A=1, X=x \right]>0$ for all $x$ and $m \in \mathcal{S}_{M_1|x}$,
\end{description}
where the supports for the mediator random variables may not be equal, and may change based on $x$.
We consider the case when $\mathcal{S}_{M_0|x}= \left\{0^* \right\}$ and
$0^* \in \mathcal{S}_{M_1|x}$ for  $X=x$ .
In the following,
 all summations are over the support $\mathcal{S}_{M_1|x}$.
\begin{eqnarray*}
E \left[ Y_{1M_0} | X=x \right] & = &  \sum_m E \left[ Y_{1m} |  M_0=m, X=x \right] Pr[ M_0=m | X=x]   \hspace*{3em} \mbox{ (by iterated expectations) } \\
& = & \sum_m E \left[ Y_{1m} | A=1, M_0=m, X=x \right] Pr[ M_0=m | A=0, X=x]  \hspace*{1em} \mbox{ (by $SI_1$) } \\
& = & \sum_m E \left[ Y_{1m} | A=1,  X=x \right] Pr[ M_0=m | A=0, X=x]    \hspace*{3em} \mbox{ (by $SI_2$) } \\
& = & \sum_m E \left[ Y_{1m} | A=1, M_1=m, X=x \right] Pr[ M_0=m | A=0, X=x]  \hspace*{1em}  \mbox{ (by $SI_2$) } \\
& = & \sum_m E \left[ Y | A=1, M=m, X=x \right] Pr[ M=m | A=0, X=x]  \hspace*{1em}  \mbox{ (by consistency) } \\
& = &  E \left[ Y | A=1, M=0^*, X=x \right]    \hspace*{4em} \mbox{ (when $Pr[ M>0^* | A=0,X=x]=0$ ) } \\
\end{eqnarray*}

\section{Proof of Theorem~1
}
\label{sec-Proof.Theorem.prsBounds}

We show the proof for $\prs$. The proof for $\pra$ is analogous and is not shown.

For $\pi$ parameters, let `$\cdot$' denote summation over an index,  so that the sums for the 5 columns under $A=0, M_0=0, Y_{00}=1$ for each of the first 3 rows of the  Table~2 are
\begin{eqnarray}
\pi_{00 \cdot 1}^{00} & = & \pi_{0001}^{00} + \pi_{0011}^{00} \label{pib} \\
\pi_{\cdot 1 \cdot 1}^{00} & = & \pi_{0101}^{00} + \pi_{0111}^{00} + \pi_{1111}^{00} \label{pic} \\
\pi_{0 \cdot \cdot 1}^{10} & = & \pi_{0001}^{10} + \pi_{0011}^{10} +  \pi_{0101}^{10} + \pi_{0111}^{10} \label{pie}
\end{eqnarray}
The issue is that we cannot differentiate among the terms in each of the sums on the right-hand side of the above equations.
Suppose we could differentiate the sum in equation~\ref{pie} such that if we knew $\tau_s$, then
\begin{eqnarray*}
 \pi_{0101}^{10} + \pi_{0111}^{10} & = & \tau_s \left( \pi_{00 \cdot 1}^{00}  + \pi_{0 \cdot \cdot 1}^{10}  \right), \mbox{ so that for any } \tau_s \in [0,1],   \\
\pi_{00 \cdot 1}^{00} + \pi_{0001}^{10} + \pi_{0011}^{10} &  = & \left( 1-\tau_s \right) \left(\pi_{00 \cdot 1}^{00} + \pi_{0 \cdot \cdot 1}^{10} \right).
\end{eqnarray*}
Since $E \left[ Y_{1 M_0}  \right] =  E \left[ Y_{1 M_1}  \right] + \pi_{0101}^{10} + \pi_{0111}^{10}$, when $\tau_s=0$ then
\begin{eqnarray*}
& E \left[ Y_{1 M_0}  \right] = & E \left[ Y_{1 M_1}  \right], \\
\Rightarrow & \theta_{\IE_s} = & \frac{E \left[ Y_{1 M_1} \right] }{E \left[ Y_{1 M_0}  \right] } = 1 \\
\Rightarrow & \prs = & \frac{ \log \left( \theta_{\IE_s}  \right) }{ \log \left( \theta_{\TE}  \right) } = 0.
\end{eqnarray*}
Since $E \left[ Y_{1 M_0}  \right] =  E \left[ Y_{0 M_0}  \right] - \left( \pi_{00 \cdot 1}^{00} + \pi_{0001}^{10} + \pi_{0011}^{10} \right)$, when $\tau_s=1$ then
\begin{eqnarray*}
& E \left[ Y_{1 M_0}  \right] = & E \left[ Y_{0 M_0}  \right], \\
\Rightarrow & \theta_{\IE_s} = & \frac{E \left[ Y_{1 M_1} \right] }{E \left[ Y_{0 M_0}  \right] } = \theta_{\TE} \\
\Rightarrow & \prs = & \frac{ \log \left( \theta_{\IE_s}  \right) }{ \log \left( \theta_{\TE}  \right) } =  1.
\end{eqnarray*}

Now we just need to show that for any values of the $\phi$ parameters (i.e., identifiable parameters) under the base model, that for any value of $\tau_s \in [0,1]$ the $\phi$ parameters do not change.
By
Table~2
we have:

\begin{tabular}{cll}
& $\phi_{vaf}  =  \pi_{1111}^{10}$                          \hspace*{2em} & $\phi_{paf}=0$ \\
& $\phi_{vnf}  =  \pi_{\cdot 1 \cdot 1}^{00}$                          \hspace*{2em} & $\phi_{pnf}  = \pi_{00 \cdot 1}^{00} + \pi_{\cdot 1 \cdot 1}^{00} +  \pi_{0 \cdot \cdot 1}^{10} + \pi_{1111}^{10}$ \\
& $\phi_{vas}  =  \pi_{0000}^{10} + \pi_{0 \cdot \cdot 1}^{10}$   \hspace*{2em} & $\phi_{pas}  = 0$ \\
& $\phi_{vns}  =  \pi_{0000}^{00} + \pi_{00 \cdot 1}^{00}$     \hspace*{2em} & $\phi_{pns}  = \pi_{0000}^{00} + \pi_{0000}^{10}$
\end{tabular}

\vspace*{3em}
If $\tau_s=0$ then $ \pi_{0101}^{10} + \pi_{0111}^{10}=0$, and we can just let $\pi_{0 \cdot \cdot 1}^{10} = \pi_{0001}^{10} + \pi_{0011}^{10}$, which does not change the $\phi$ parameters. If $\tau_s=1$
then $\pi_{00 \cdot 1}^{00} + \pi_{0001}^{10} + \pi_{0011}^{10}=0$, and it is straightforward to allow $\pi_{0001}^{10} + \pi_{0011}^{10}=0$, but setting $\pi_{00 \cdot 1}^{00}$ (equation~\ref{pib}) to $0$ can also be accomplished without changing the $\phi$ parameters.
If $\pi_{00 \cdot 1}^{00}=0$ then $\pi_{0000}^{00}=\phi_{vns}$, so that $\pi_{0000}^{10}=\phi_{pns}-\phi_{vns}$, and $\pi_{0 \cdot \cdot 1}^{10}=\phi_{vas} - \phi_{pns} + \phi_{vns} = \phi_{vs} - \phi_{pns}$, and neither of $\pi_{\cdot 1 \cdot 1}^{00}$ and $\pi_{1111}^{10}$ will change; therefore,
we can set $\pi_{00 \cdot 1}^{00}=0$  without changing any $\phi$ parameter. Because the bounds $\tau_s=0$ and $\tau_s=1$ do not change the $\phi$ parameters, by continuity, all $\tau_s \in (0,1)$ also do not change the $\phi$ parameters.

\section{Proof of Theorem~2
}
\label{app-Proof.Theorem.MindY}

Consider first statement (i).
Let $[ \pi_{0000}, \pi_{0001}, \pi_{0011}, \pi_{0101}, \pi_{0111}, \pi_{1111}]$ be the marginal probability for the 6 possible potential outcomes, ${\bf Y}=[Y_{11},Y_{10},Y_{01},Y_{00}]$,
from Table~2, which sum to $1$.  Let $\phi_{vn} = \phi_{vnf}+\phi_{vns}$ be $Pr[ M_1=0]$,
and let $\phi_{va} = \phi_{vaf}+\phi_{vas}$ be $Pr[ M_1=1]=1-Pr[M_1=0]$.
Under $Y_{am} \ind M_{a'}$ for all $a,a',m$,
we have:
\begin{eqnarray}
\phi_{vn}
\left[ \begin{array}{c}
\pi_{0000} \\
\pi_{0001} \\
\pi_{0011} \\
\pi_{0101} \\
\pi_{0111} \\
\pi_{1111}
\end{array}
\right] & = &
\left[ \begin{array}{c}
\pi_{0000}^{00} \\
\pi_{0001}^{00} \\
\pi_{0011}^{00} \\
\pi_{0101}^{00} \\
\pi_{0111}^{00} \\
\pi_{1111}^{00}
\end{array}
\right]   \label{eq:omega1}
\end{eqnarray}

and
\begin{eqnarray}
\phi_{va}
\left[ \begin{array}{c}
\pi_{0000} \\
\pi_{0001} \\
\pi_{0011} \\
\pi_{0101} \\
\pi_{0111} \\
\pi_{1111}
\end{array}
\right] & = &
\left[ \begin{array}{c}
\pi_{0000}^{10} \\
\pi_{0001}^{10} \\
\pi_{0011}^{10} \\
\pi_{0101}^{10} \\
\pi_{0111}^{10} \\
\pi_{1111}^{10}
\end{array}
\right] \label{eq:omega2}
\end{eqnarray}

Using the first rows of equations~\ref{eq:omega1} and \ref{eq:omega2}, and the fact that $\phi_{pns} = \pi_{0000}^{00} + \pi_{0000}^{10}$, we get
\begin{eqnarray*}
& & \phi_{pns} = \pi_{0000} \phi_{vn} + \pi_{0000} \phi_{va}  = \pi_{0000}  \\
\Rightarrow & & \pi_{0000}^{00} = \phi_{pns} \phi_{vn}  \\
\Rightarrow & & \pi_{0000}^{10} = \phi_{pns} \phi_{va}
\end{eqnarray*}
Then since $\phi_{vns} =\pi_{0000}^{00}+ \pi_{00 \cdot 1}^{00}$, we have
\begin{eqnarray}
 \pi_{00 \cdot 1}^{00}= \phi_{vns} - \phi_{pns} \phi_{vn}, \label{eq:pib.ZindY}
\end{eqnarray}
and since $\phi_{vas} = \pi_{0000}^{10} + \pi_{0 \cdot \cdot 1}^{10}$, we have
\begin{eqnarray}
\pi_{0 \cdot \cdot 1}^{10}= \phi_{vas} - \phi_{pns} \phi_{va}. \label{eq:pie.ZindY}
\end{eqnarray}
From Tables~1 and 2
we get $\pi_{\cdot 1 \cdot 1}^{00}=\phi_{vnf}$ and $\pi_{1111}^{10}=\phi_{vaf}$, and this completes the proof of statement (i).

For statement (iii), we can similarly use the last rows  of equations~\ref{eq:omega1} and \ref{eq:omega2} to solve for $\pi_{1111}^{00}$.
We have $\pi_{1111} = \pi_{1111}^{00} / \phi_{vn} = \pi_{1111}^{10} / \phi_{va} = \phi_{vaf} / \phi_{va}$ giving
$\pi_{1111}^{00} = \phi_{vaf} \phi_{vn}/\phi_{va}$. Then
\begin{eqnarray*}
\pi_{0101}^{00} + \pi_{0111}^{00} = \pi_{\cdot 1 \cdot 1}^{00} - \pi_{1111}^{00} =
\phi_{vnf} - \phi_{vaf} \phi_{vn}/\phi_{va}. \label{eq:c1+c2}
\end{eqnarray*}
Then using the fourth and fifth rows of equations~\ref{eq:omega1} and \ref{eq:omega2}, we get
$\pi_{0101} + \pi_{0111} =  \left( \pi_{0101}^{00} + \pi_{0111}^{00} \right)/\phi_{vn} =  \left( \pi_{0101}^{10} + \pi_{0111}^{10} \right)/\phi_{va}$
giving
\begin{eqnarray*}
\pi_{0101}^{10} + \pi_{0111}^{10} & = & \frac{  \phi_{va} }{\phi_{vn} } \left( \phi_{vnf} - \phi_{vaf} \phi_{vn}/\phi_{va} \right)  \\
& = & \frac{ \phi_{va} \phi_{vnf} }{ \phi_{vn} } - \phi_{vaf}.
\end{eqnarray*}
Thus,
\begin{eqnarray*}
E \left[ Y_{1M_0} \right] &  =  & E \left[ Y_{1M_1} \right] +\pi_{0101}^{10} + \pi_{0111}^{10} \\
& = & \phi_{vaf} + \phi_{vnf} + \frac{ \phi_{va} \phi_{vnf} }{ \phi_{vn} } - \phi_{vaf} \\
& = & \phi_{vnf} \left( 1 + \frac{ \phi_{va}  }{ \phi_{vn} } \right)  = \frac{ \phi_{vnf} }{ \phi_{vn} }
\end{eqnarray*}
where in the last line we use $\phi_{va}+\phi_{vn}=1$. This proves statement (iii).

To show statement (ii), that $E \left[ Y_{0M_1} \right]$ is not identifiable under $\mathcal{M}_{2}$, we note that
 $E \left[ Y_{0M_1} \right] = E \left[ Y_{0M_0} \right] - \left( \pi_{0001}^{10} + \pi_{0101}^{10} \right)$.
 We have shown that $\pi_{0101}^{10}+\pi_{0111}^{10}$  are identifiable, and $\pi_{0 \cdot \cdot 1}^{10}$ is identifiable;
therefore,  $\pi_{0001}^{10}+\pi_{0011}^{10}$ is identifiable. In order  to differentiate between $\pi_{0001}^{10}$ and $\pi_{0011}^{10}$
we would need to be able to either (i) differentiate the potential outcome vectors between individuals with ${\bf Y}=[0001], {\bf M}=[10]$ and individuals with  ${\bf Y}=[0011], {\bf M}=[10]$,
or (ii) differentiate the potential outcome vectors between  with ${\bf Y}=[0001], {\bf M}=[00]$ and individuals with  ${\bf Y}=[0011], {\bf M}=[00]$, which have the same potential response outcome vectors as in (i),
but different values of the potential mediator responses.
From Table~2 there is no way to differentiate between individuals with ${\bf Y}=[0001]$ and individuals with ${\bf Y}=[0011]$, regardless of the value of ${\bf M}$,
and hence  $E \left[ Y_{0M_1} \right]$ is not identifiable, and statement (ii) is proved.

\section{Proof of Theorem~3
}
\label{sec-Proof.Theorem.EquivProblems}

The conditions of Statement~A can be found by writing the expectations in terms of the $\pi$ parameters (see Table~2),
\begin{eqnarray*}
E(Y_{1M_1}) = \pi_{\cdot 1 \cdot 1}^{00} + \pi_{1111}^{10} \leq
E(Y_{1M_0}) = E(Y_{1M_1}) + \pi_{0101}^{10}+\pi_{0111}^{10} \leq
E(Y_{0M_0}) = E(Y_{1M_0}) + \pi_{00 \cdot 1} + \pi_{0001}^{10} + \pi_{0011}^{10},
\end{eqnarray*}
where $\pi_{\cdot 1 \cdot 1}^{00} = \pi_{0101}^{00} + \pi_{0111}^{00} + \pi_{1111}^{00}=\phi_{vnf}$. The result in Statement A is Theorem~2(iii).

Statement~B immediately follows from writing
\begin{eqnarray*}
\prs = \frac{ \log \left\{ E(Y_{1M_1}) \right\} -  \log \left\{ E(Y_{1M_0}) \right\} }{ \log \left\{ E(Y_{1M_1}) \right\} -  \log \left\{ E(Y_{0M_0}) \right\} }.
\end{eqnarray*}

For Statement~C, we first note that $Y_{am} \ind M_{a'}$ implies that $M_1 \ind Y_{10}$,
and since $M_0=0$ for all individuals in the base model, $Y_{1M_0}=Y_{10}$ (see also Table~1).
By definition, the correlation between $M_1$ and $Y_{1M_0}$ is
\begin{eqnarray}
\mathrm{Corr} \left( M_1, Y_{1M_0} \right) & = & \frac{ E \left( M_1  Y_{1M_0} \right) - E \left( M_1 \right) E \left( Y_{1M_0} \right) }{ \sqrt{ \mathrm{Var} \left( M_1 \right)  \mathrm{Var} \left( Y_{1M_0}  \right) }}  \nonumber \\
& = & \frac{ \left\{  E \left( Y_{1M_0} \right) - E \left( Y_{1M_1} | M_1=0 \right) \right\} Pr \left[ M_1=0 \right] }{
 \sqrt{ E \left( M_1 \right) \left\{ 1 - E \left( M_1 \right) \right\}  E \left( Y_{1M_0}  \right) \left\{ 1 - E \left(Y_{1M_0}  \right) \right\} }}  \label{eq:CorrM1.Y1M0.orig}
\end{eqnarray}
where the last step uses $Y_{1M_0}=Y_{10}$ since $Pr[M_0=0]=1$, and
\begin{eqnarray*}
E \left( M_1  Y_{1M_0} \right)  - E \left( M_1 \right) E \left( Y_{1M_0} \right) & = & \frac{ Pr[ Y_{10}=1 \;\; \& \;\; M_1=1]}{ Pr[ M_1=1] }  Pr[ M_1=1] - E \left( M_1 \right) E \left( Y_{10} \right)  \\
& = &  Pr \left[ Y_{10}=1 | M_1=1 \right] Pr[ M_1=1] - Pr \left( M_1=1 \right) E \left( Y_{10} \right) \\
 & = &  Pr \left[ Y_{10}=1  \right]  -  Pr \left[ Y_{10}=1 | M_1=0 \right] Pr[ M_1=0] - Pr \left( M_1=1 \right) E \left( Y_{10}  \right) \\
 & = &  E \left[ Y_{10} \right] - Pr \left( M_1=1 \right) E \left( Y_{10}  \right)  -  Pr \left[ Y_{10}=1 | M_1=0 \right] Pr[ M_1=0]  \\
  & = &  E \left[ Y_{10} \right] \left\{ 1 - Pr \left( M_1=1 \right) \right\}  -  Pr \left[ Y_{10}=1 | M_1=0 \right] Pr[ M_1=0]  \\
 & = & \left\{  E \left[ Y_{10} \right] - E \left[ Y_{10} | M_1=0 \right] \right\} Pr[ M_1=0]  \\
  & = & \left\{  E \left[ Y_{1M_0} \right] - E \left[ Y_{1M_1} | M_1=0 \right] \right\} Pr[ M_1=0]  \\
  & = &  \left\{  E \left[ Y_{1M_0} \right] - \frac{\phi_{vnf}}{\phi_{vn}}  \right\} \phi_{vn}  \\
\end{eqnarray*}
which is a function of $E \left( Y_{1M_0} \right)$ and identifiable parameters.
Substituting $\phi_{vn}$ for $1-E(M_1)$, we get
\begin{eqnarray}
\mathrm{Corr} \left( M_1, Y_{1M_0} \right) & = & \frac{ \left\{  E \left( Y_{1M_0} \right) - \frac{\phi_{vnf}}{\phi_{vn}} \right\} \phi_{vn} }{
 \sqrt{ (1- \phi_{vn} ) \phi_{vn}   E \left( Y_{1M_0}  \right) \left\{ 1 - E \left(Y_{1M_0}  \right) \right\} }}  \label{eq:CorrM1.Y1M0}
\end{eqnarray}
Substituting $E[Y_{1M_0}]= \frac{ \phi_{vnf}}{\phi_{vn}}$ gives $\mathrm{Corr} \left( M_1, Y_{1M_0} \right) = 0$.
From Section~\ref{sec-Proof.Theorem.prsBounds},
\begin{eqnarray*}
\phi_{vf} =  E \left( Y_{1M_1} \right) & \leq & E \left( Y_{1M_0} \right) \leq E \left( Y_{0M_0} \right) = \phi_{pnf}.
\end{eqnarray*}
Substituting the minimum and maximum values for $E \left( Y_{1M_0} \right)$ in terms of $\phi$ parameters gives $\rho_{min}$ and $\rho_{max}$, respectively.


\end{document}